\documentclass[twocolumn,amsmath,amssymb,
aps, showpacs,graphicx]{revtex4-2}

\usepackage{graphicx}
\usepackage{dcolumn}
\usepackage{bm}
\usepackage{lineno, hyperref}
\usepackage{natbib}
\begin{document}

\title{High thermoelectric performance in excitonic bilayer graphene}

\author{V. Apinyan} 
\altaffiliation[e-mail:]{v.apinyan@intibs.pl}
\author{T. K. Kope\'{c}} 
\affiliation{Institute of Low Temperature and Structure Research, Polish Academy of Sciences\\ PO. Box 1410, 50-950 Wroc\l{}aw 2, Poland}
%

\begin{abstract}
We consider the excitonic effects on the thermal properties in the AB-stacked bilayer graphene. The calculations are based on the bilayer generalization of the usual Hubbard model at the half-filling. The full interaction bandwidth is used without any low-energy assumption. We obtain the unusually high values for the electronic figure of merit even at the room-temperatures which is very promising for the thermoelectric applications of the AB-bilayer structure. We discuss the effects of the interlayer Coulomb interaction and temperature on different thermal parameters in the bilayer graphene and we emphasize the role of the charge neutrality point in the thermal properties and within the excitonic insulator transition scenario. The calculated values of the rate of thermoelectric conversion efficiency suggest the possibility of high-performance device applications of AB-bilayer graphene. 
\end{abstract}



\maketitle

\section{\label{sec:Section_1} Introduction}
%
The discovery of the very high heat conductivity in graphene has led to a great scientific interest and surges many experimental and theoretical studies \cite{cite_1,cite_2,cite_3,cite_4, cite_5, cite_6,cite_7,cite_8, cite_9} on the thermal and thermoelectric properties of this unusual material. The
anomalous high values of thermopower have been observed also in graphene monolayer in contact with a semiconductor substrate \cite{cite_10,cite_11}. Recently, the thermal properties in bilayer graphene (BLG) systems have been also under intense studies \cite{cite_12,cite_13,cite_14,cite_15,cite_16, cite_17, cite_18, cite_19, cite_20} owing to the plausible device applications of the bilayer graphene as well as fundamental interest in this material due to the large band gap opening comparable to those in semiconducting systems \cite{cite_21}. The fundamental effect which observed experimentally was the electric field tunability of the thermoelectric power (TEP) \cite{cite_14} under the applied perpendicular electric field which opened new possibilities for the thermoelectric applications using bilayer graphene-based devices. Recently, it has been predicted, theoretically, \cite{cite_19} that the electric field effect could enhance the room-temperature thermopower in bilayer graphene more than four times that of the monolayer graphene and unbiased bilayer graphene. This effect has been confirmed by the experimental results at the low-temperature and with the dual gated bilayer graphene \cite{cite_22}. It has been shown that one can realize even higher thermoelectric power in a clean dual-gated bilayer graphene device at the large electrical displacement field. The frequency dependence of the thermal conductivity in the AA and AB bilayer graphene and also the bias voltage dependence have been intensively studied in a series of theoretical works \cite{cite_23, cite_24, cite_25, cite_26}. The thermoelectric efficiency of a high-performance thermoelectric material, for both power generation and cooling, is determined by its own enhancing (dimensionless) figure of merit (FOM) $ZT$, discussed well in Ref. \cite{cite_27}, where a $\delta$-shaped transport distribution is found to maximize the thermoelectric properties. It has been shown in Ref.\cite{cite_28} that the thermoelectric power in bilayer graphene can be drastically increased by introducing the nanopores and by shifting the Fermi level. In this way, the authors obtained for the electronic part of the thermoelectric figure of merit the room temperature value $ZT=2.45$ which renders the bilayer graphene as to be competitive with other power generation systems. It has been shown recently \cite{cite_29} that the thermoelectric figure of merit is considerably enhanced in bilayer graphene at the low-temperature limit ($ZT>3$) when a rigid substrate suppresses out-of-plane phonons by reducing their contribution to the thermal conductivity. By using an atomic tight-binding Hamiltonian, the authors in Refs.\cite{cite_30, cite_31,cite_32} have shown that one can achieve very high values of $ZT$ in graphene devices composed of two partially overlapped graphene sheets coupled to each other in the cross-plane direction. The authors have shown that the figure of merit can reach the values higher than one in the junctions consisting especially of gapped graphene materials. Particularly, they have revealed that the phonon conductance in these layered structures gets strongly reduced compared to that in single layer graphene. The excitons can also be involved in thermal conductivity. The excitonic thermal conductivity has been reported in a few works \cite{cite_33, cite_34, cite_35}. For the first time, the excitonic thermal conductivity was calculated in Ref.\cite{cite_33}. The thermal conductivity of the excitonic insulator has been calculated in Ref.\cite{cite_36}, in the semimetallic limit. Experimentally, the excitonic thermal conductivity was measured only in 1988 and reported in Ref.\cite{cite_37}, where the provided data evidently shows the possibility of excitonic thermal transport in solids. 
Moreover, it appears that the bilayer exciton system displays a remarkable similarity to a thermoelectric couple as it was suggested in Ref.\cite{cite_38}, where the authors suggested the experimental realization of a thermocouple device based on the bilayer-excitons, which has the potential to increase considerably the figure of merit ZT. 
Recently, the thermal transport in the exciton-condensate Josephson junction, composed of two graphene bilayers, has been studied \cite{cite_39} and the quasiparticle and interference heat currents through an insulating barrier have been calculated. It was shown in Ref.\cite{cite_39} that the phase difference of the condensates has an important impact on the thermal rectification effect in the junction. 

Notwithstanding the numerous studies on the thermal properties in bilayer graphene heterostructures, there is no information about the influence and the role of the electron-hole bonding and electron-electron interaction on the thermal properties in the single bilayer graphene. In the present paper, we decided to accomplish this lack in the literature and we considered the excitonic effects in the thermal properties of the bilayer graphene structure. Basing on our previous results, concerning the excitonic gap parameter and chemical potential \cite{cite_38} in the suspended AB stacked bilayer graphene with excitons, we considered here the interacting bilayer Hubbard model with the full bandwidth of the interaction strength and without any low-energy assumptions. By using the equations of motion for the electric and heat current densities in the system and by employing the Kubo linear response theory, we calculated the in-plane components of the heat transport coefficients and we found their temperature dependence. The calculations show that the excitonic effects strongly enhance the thermoelectric figure of merit ($2.5<ZT_{\rm e}<3$ at the room temperatures) for a given range of frequencies of the external photonic field and for the selected values of the local interlayer interaction parameter. Moreover, the high values of the thermoelectric conversion efficiency parameter indicate about the important role that play the excitons in AB-BLG which opens the possibility for the direct device application of the bilayer graphene as a promising thermoelectric material. Particularly, the thermoelectric energy conversion efficiency is strongly enhanced ($>20 \%$) at the high photon's frequencies and at the large Carnot efficiency. A detailed discussion about all obtained thermoelectric effects is given and the conditions for the high thermoelectric performance in AB-BLG system are highlighted. The presented work could have the important impact in the technological applications of the bilayer graphene as an efficient thermoelectric generator, which operates at the room temperatures. 

The paper is organized as follows: in the Section \ref{sec:Section_2}, we introduce the considered model and we discuss the single-particle and excitonic Green's functions. In the Section \ref{sec:Section_3}, we calculate the electronic and thermal current densities within the Mahan's formalism and in Section \ref{sec:Section_4} we present the thermal kinetic coefficients within the Kubo theory. Furthermore, in the Section \ref{sec:Section_5}, we give the numerical calculation results and in Section \ref{sec:Section_6} we give a conclusion to our paper. In the \ref{sec:Section_7}, the electronic and heat current density operators are calculated in details and \ref{sec:Section_8} is devoted to the calculation of heat-heat response function. 
%
\section{\label{sec:Section_2} The bilayer Hubbard model}
%
\begin{figure}
	\begin{center}
		\includegraphics[scale=0.6]{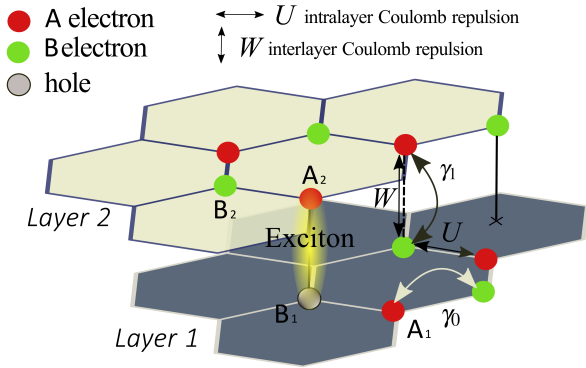}
		\caption{\label{fig:Fig_1}(Color online) The schematic picture of the AB-stacked bilayer graphene system. Different physical parameters are shown in the picture which enter into the expression of the Hamiltonian of the system, given in Eq.(\ref{Equation_1}). The excitonic formation is shown in the picture in the form of the light-yellow cloud.}
	\end{center}
\end{figure} 
%
Here, we consider the AB Bernal stacked bilayer graphene (BLG) and we describe it with the help of the interacting bilayer Hubbard Hamiltonian. The AB-BLG system is composed of two coupled honeycomb layers with sublattices $A_{1}$, $B_{1}$ and ${A}_{2}$, ${B}_{2}$, in the bottom and top layers respectively. In the $z$-direction the layers are arranged according to Bernal Stacking (BS) order \cite{cite_39}, i.e. the atoms on sites ${A}_{2}$ in the top layer lie just above the atoms on sites $B_{1}$ in the bottom layer graphene, and each layer is composed of two inter-penetrating triangular lattices. The schematic representation of the AB-stacked bilayer graphene is shown in Fig.~\ref{fig:Fig_1}. The Hamiltonian of the interacting BLG system at the half-filling is 
\begin{eqnarray}
   	&&H=-\gamma_0\sum_{\left\langle {\bf{m}}{\bf{m}}'\right\rangle}\sum_{\ell\sigma}\left({a}^{\dag}_{\ell,{\bf{m}}\sigma}b_{\ell,{\bf{m}}'\sigma}+{\rm h.c.}\right)
   	\nonumber\\
   	&&-\gamma_1\sum_{{\bf{m}}\sigma}\left({{b}}^{\dag}_{1,{\bf{m}}\sigma}{a}_{2,{\bf{m}}\sigma}+{\rm h.c.}\right)-\sum_{\ell=1,2}\sum_{{\bf{m}}\eta}\mu_{\ell}n_{\ell,\eta{\bf{m}}}
   	\nonumber\\
   	&&+U\sum_{{\bf{m}}}\sum_{\ell\eta}\left[\left(n_{\ell,\eta{\bf{m}}\uparrow}-1/2\right)\left(n_{\ell,\eta{\bf{m}}\downarrow}-1/2\right)-1/4\right]
   	\nonumber\\
   	&&+W\sum_{{\bf{m}}\sigma\sigma'}\left[\left(n_{b_1,{\bf{m}}\sigma}-1/2\right)\left(n_{{a_2},{\bf{m}}\sigma'}-1/2\right)-1/4\right],
   	\label{Equation_1}
   	\end{eqnarray}
where the parameters $\gamma_{0}$ and $\gamma_{1}$ are the intralayer and interlayer hopping amplitudes. The summation $\left\langle {\bf{m}}{\bf{m}}' \right\rangle$, in the first term, in Eq.(\ref{Equation_1}), denotes the sum over the nearest neighbors lattice sites ${\bf{m}}$ and ${\bf{m}}'$ in different honeycomb layers in the bilayer graphene. We keep the small letters $a_1,b_1$ and ${a}_2,{b}_2$ for the electron operators on the lattice sites $A_{1},B_{1}$ and ${A}_{2},{B}_{2}$ respectively.
The summation index $\ell=1,2$ denotes the layers in the BLG. Particularly, we use $\ell=1$ for the bottom layer, and $\ell=2$ for the top layer. Furthermore, we have $\eta=a_1, b_1$ for $\ell=1$, and $\eta={a}_2, {b}_2$, for $\ell=2$. The spin variable $\sigma$ takes two possible directions ($\sigma= \uparrow, \downarrow$). Next, $n_{\ell,\eta{\bf{m}}}$ in Eq.(\ref{Equation_1}) is the total electron density operator for the layer $\ell$
\begin{eqnarray}
n_{\ell,\eta{\bf{m}}}=\sum_{\sigma=\uparrow,\downarrow}n_{\ell,\eta{\bf{m}}\sigma},
\label{Equation_2}
\end{eqnarray}
and $n_{\ell,{\bf{m}}\eta\sigma}={\eta}_{\ell,{\bf{m}}\sigma}\eta_{\ell,{\bf{m}}\sigma}$ is the electron density operator for the $\eta$-type fermion in the layer $\ell$ and with the attached spin $\sigma$.
We consider the suspended BLG structure with pure electronic layers and without initial doping in the system. The condition of half-filling in each layer reads as $\left\langle n_{\ell} \right\rangle=1$, for $\ell=1,2$, where $n_{\ell}$ is the total electron density operator for the layer $\ell$. Furthermore, $U$, in the Hubbard term in Eq.(\ref{Equation_1}), parametrizes the intralayer Coulomb interaction, while the parameter $W$ denotes the local interlayer Coulomb repulsion between the electrons located on sites $B_{1}$ and ${A}_{2}$, in different layers. In Fig.~\ref{fig:Fig_1}, here, we have shown different physical parameters entering in the Hamiltonian, in Eq.(\ref{Equation_1}). The excitonic formation between the layers is also shown in Fig.~\ref{fig:Fig_1}. We will put $\gamma_{0}=1$, as the unit of energy in the system, and we set $k_{B}=1$, $\hbar=1$ and $e=1$ throughout the paper. For a simple treatment at equilibrium, we initially suppose the balanced BLG structure, i.e., with the equal chemical potentials in the both layers $\mu_1=\mu_2$. The Hamiltonian, given in Eq.(\ref{Equation_1}) could be reduced to an effective Hamiltonian which is linear in the electron density terms. This procedure is discussed in details in Ref.\cite{cite_38} and we give in \ref{sec:Section_7} the linearized form of it after the Hubbard-Stratanovich decouplings of the interaction terms.
%
\subsection{\label{sec:Section_2_1} The partition function and the Green's functions in the BLG}
%
We write the total action of the BLG system in the following form 
\begin{eqnarray}
&&S\left[{a}^{\dag},a,{b}^{\dag},b\right]=\sum_{\ell=1,2}S^{(\ell)}_{\rm B}\left[{a}^{\dag}_{\ell},a_{\ell}\right]+\sum_{\ell=1,2}S^{(\ell)}_{\rm B}\left[{b}^{\dag}_{\ell},b_{\ell}\right]
\nonumber\\
&&+\int^{\beta}_{0}d\tau H\left(\tau\right).
\label{Equation_3}
\end{eqnarray}
Here, the first two terms are the Berry terms for the layers with the indices $\ell=1,2$
\begin{eqnarray}
S^{(l)}_{\rm B}\left[{a}^{\dag}_{\ell},a_{\ell}\right]=\sum_{{\bf{m}},\sigma}\int^{\beta}_{0}d\tau {a}^{\dag}_{l,{\bf{m}}\sigma}(\tau)\frac{\partial}{\partial \tau}a_{l,{\bf{m}}\sigma}(\tau),
\nonumber\\
S^{(l)}_{\rm B}\left[{b}^{\dag}_{\ell},b_{\ell}\right]=\sum_{{\bf{m}},\sigma}\int^{\beta}_{0}d\tau {b}^{\dag}_{l,{\bf{m}}\sigma}(\tau)\frac{\partial}{\partial \tau}b_{l,{\bf{m}}\sigma}(\tau),
\label{Equation_4}
\end{eqnarray}
where we have introduced the imaginary time $\tau$ at each lattice site position ${\bf{m}}$. The upper limit of integration over $\tau$, in all terms in Eq.(\ref{Equation_3}), is given as $\beta=1/T$ where $T$ is the temperature. The Hamiltonian $H\left(\tau\right)$ of the BLG system, in the last term in Eq.(\ref{Equation_3}), is described in Eq.(\ref{Equation_1}). Furthermore, after some linearization procedure \cite{cite_38, cite_40} of the nonlinear interaction terms in Eq.(\ref{Equation_1}) and transforming the fermionic operators into the Fourier space representation we can rewrite the fermionic action in the following form
\begin{eqnarray} 
S\left[{\Psi}^{\dag},\Psi\right]=\frac{1}{\beta{N}}\sum_{{\bf{k}}\nu_{n}, \sigma}{\Psi}^{\dag}_{{\bf{k}},\sigma}(\nu_{n}){\cal{G}}^{-1}_{{\bf{k}},\sigma}(\nu_{n}){\Psi}_{{\bf{k}},\sigma}(\nu_{n}),
\label{Equation_5}
\end{eqnarray}
where we have introduced the four component fermionic Nambu-spinors ${\Psi}_{{\bf{k}},\sigma}(\nu_{n})=\left[a_{1{\bf{k}},\sigma},b_{1{\bf{k}},\sigma},{a}_{2{\bf{k}},\sigma},{b}_{2{\bf{k}},\sigma}\right]^{T}$ at each discrete state ${\bf{k}}$ in the reciprocal space and for each spin direction $\sigma=\uparrow, \downarrow$. The fermionic Matsubara frequencies $\nu_{n}$ are $\nu_{n}=\pi(2n+1)/\beta$ with $n=0,\pm 1, \pm 2 ...$.  Next, ${\cal{G}}^{-1}_{{\bf{k}},\sigma}(\nu_{n})$ stands for a $4\times 4$ inverse Green's function matrix. It is defined as follows
\begin{eqnarray}
{\cal{G}}^{-1}_{{\bf{k}},\sigma}\left(\nu_{n}\right)=\left(
\begin{array}{ccccrrrr}
E_{1}(\nu_{n}) & -\tilde{\gamma}_{1{\bf{k}}} & 0 & 0\\
-\tilde{\gamma}^{\ast}_{1{\bf{k}}} &E_{2}(\nu_{n})  & -\gamma_{1}-{\Delta}^{\dag} & 0 \\
0 & -\gamma_{1}-{\Delta} & E_{2}(\nu_{n}) & -\tilde{\gamma}_{2{\bf{k}}} \\
0 & 0 & -\tilde{\gamma}^{\ast}_{2{\bf{k}}} & E_{1}(\nu_{n}) 
\end{array}
\right).
\nonumber\\
\label{Equation_6}
\end{eqnarray}
Here, the diagonal elements in the matrix ${\cal{G}}^{-1}_{{\bf{k}},\sigma}$ are defined as $E_{1}(\nu_{n})=-i\nu_{n}-\mu^{\rm eff}_{1}$ and $E_{2}(\nu_{n})=-i\nu_{n}-\mu^{\rm eff}_{2}$, where the effective chemical potentials $\mu^{\rm eff}_{1}$ and $\mu^{\rm eff}_{2}$, have been introduced in the form $\mu^{\rm eff}_{1}=\mu+U/4$, $\mu^{\rm eff}_{2}=\mu+U/4+W$. The parameter $\Delta$ in Eq.(\ref{Equation_6}) represents the excitonic order parameter which is defined as $\Delta=W\left\langle {b}^{\dag}_{1,\sigma}({\bf{r}},\tau){a}_{2,\sigma}({\bf{r}},\tau)\right\rangle$. We considered here the case of the homogeneous BLG structure where the pairing is between the particles with the same spin orientations, i.e. $\Delta_{\sigma\sigma'}=\Delta_{\sigma}\delta_{\sigma\sigma'}\equiv \Delta$.
The parameter $\tilde{\gamma}_{\ell{\bf{k}}}$, in Eq.(\ref{Equation_6}), is the renormalized energy dispersion in the layer $\ell$ and $\tilde{\gamma}_{\ell{\bf{k}}}=\gamma_{\ell{\bf{k}}}t$, where $\gamma_{\ell{\bf{k}}}=\sum_{\bm{\mathit{\delta}}}e^{-i{{\bf{k}}\bm{\mathit{\delta}}}_{\ell}}$. The vectors $\bm{\mathit{\delta}}_{\ell}$ are the nearest neighbor vectors in different layers $\ell=1,2$. The components of $\bm{\mathit{\delta}}_{\ell}$, for the layer with $\ell=1$, are $\bm{\mathit{\delta}}^{(1)}_{1}=\left({a_{0}}/{2\sqrt{3}},a_{0}/2\right)$, $\bm{\mathit{\delta}}^{(2)}_{1}=\left({a_{0}}/{2\sqrt{3}},-a_{0}/2\right)$, $\bm{\mathit{\delta}}^{(3)}_{1}=\left(-a_{0}/\sqrt{3},0\right)$, and $a_{0}=\sqrt{3}a$ is the sublattice constant, while $a$ is the carbon-carbon length in the graphene sheets). In the layer with $\ell=2$ (the top layer in the stacking), we have $\bm{\mathit{\delta}}^{(1)}_{2}=\left(a_{0}/\sqrt{3},0\right)$, $\bm{\mathit{\delta}}^{(2)}_{2}=\left(-{a_{0}}/{2\sqrt{3}},-a_{0}/2\right)$, $\bm{\mathit{\delta}}^{(3)}_{2}=\left(-{a_{0}}/{2\sqrt{3}},a_{0}/2\right)$. It is not difficult to realise that $\bm{\mathit{\delta}}_{2}=-\bm{\mathit{\delta}}_{1}$. For the function $\gamma_{1{\bf{k}}}$, we have $\gamma_{1{\bf{k}}}=e^{-ik_{x}a}+2e^{i\frac{k_{x}a}{2}}\cos{\frac{\sqrt{3}}{2}k_{y}a}$. By the convention, we put $a\equiv 1$, for both layers. It is not difficult to realize that $\gamma_{2{\bf{k}}}=\gamma^{\ast}_{1{\bf{k}}}\equiv\gamma^{\ast}_{{\bf{k}}}$ and it follows that $\tilde{\gamma}_{2{\bf{k}}}=\tilde{\gamma}^{\ast}_{1{\bf{k}}}\equiv\tilde{\gamma}^{\ast}_{{\bf{k}}}$, where we have omitted the layer index $\ell$. 
The partition function of the system takes the following form
\begin{eqnarray} 
{\cal{Z}}=\int{\left[{\cal{D}}\Psi^{\dag}{\cal{D}}\Psi\right]}e^{-S\left[{\Psi}^{\dag},\Psi\right]},
\label{Equation_7}
\end{eqnarray}
where the action in the exponent is given in Eq.(\ref{Equation_5}). 
Furthermore, by assuming the half-filling condition in each layer of BLG, we obtain a set of self-consistent equations for the excitonic order parameter $\Delta$ and the chemical potential $\mu$ in the BLG 
\begin{eqnarray}
&&\frac{4}{N}\sum_{{\bf{k}}}\sum^{4}_{i=1}\alpha_{i{{\bf{k}}}}n_{\rm F}(\mu-\varepsilon_{i{\bf{k}}})=1,
\label{Equation_8}
\newline\\
&&\Delta=\frac{W(\gamma_{1}+\Delta)}{N}\sum_{{\bf{k}}}\sum^{4}_{i=1}\beta_{i{{\bf{k}}}}n_{\rm F}(\mu-\varepsilon_{i{\bf{k}}}),
\label{Equation_9}
\end{eqnarray}
where the dimensionless coefficients $\alpha_{i{{\bf{k}}}}$, in Eq.(\ref{Equation_8}) with $i=1,..4$, are given as
\begin{eqnarray}
\footnotesize
\arraycolsep=0pt
\medmuskip = 0mu
\alpha_{i{{\bf{k}}}}
=(-1)^{i+1}
\left\{
\begin{array}{cc}
\displaystyle  & \frac{{\cal{P}}^{(3)}(\varepsilon_{i{\bf{k}}})}{\left(\varepsilon_{1{\bf{k}}}-\varepsilon_{2{\bf{k}}}\right)}\prod^{}_{j=3,4}\frac{1}{\left(\varepsilon_{i{\bf{k}}}-\varepsilon_{j{\bf{k}}}\right)},  \ \ \  $if$ \ \ \ i=1,2,
\newline\\
\newline\\
\displaystyle  & \frac{{\cal{P}}^{(3)}(\varepsilon_{i{\bf{k}}})}{\left(\varepsilon_{3{\bf{k}}}-\varepsilon_{4{\bf{k}}}\right)}\prod^{}_{j=1,2}\frac{1}{\left(\varepsilon_{i{\bf{k}}}-\varepsilon_{j{\bf{k}}}\right)}.  \ \ \  $if$ \ \ \ i=3,4.
\end{array}\right.
\nonumber\\
\label{Equation_10}
\end{eqnarray}
Here, ${\cal{P}}^{(3)}(x)$ is a polynomial of third order ${\cal{P}}^{(3)}(x)=x^{3}+a_{1{\bf{k}}}x^{2}+a_{2{\bf{k}}}x+a_{3\bf{k}}$ with the coefficients 
\begin{eqnarray}
&&a_{1{\bf{k}}}=-2\mu^{\rm eff}_{2}-\mu^{\rm eff}_{1},
\nonumber\\
&&a_{2{\bf{k}}}=\mu^{\rm eff}_{1}\left(\mu^{\rm eff}_{2}+2\mu^{\rm eff}_{1}\right)-(\Delta+\gamma_1)^{2}-|\tilde{\gamma}_{{\bf{k}}}|^{2}, 
\nonumber\\
&&a_{3{\bf{k}}}=-\mu^{\rm eff}_{1}\left(\mu^{\rm eff}_{2}\right)^{2}+\mu^{\rm eff}_{1}(\Delta+\gamma_1)^{2}+\mu^{\rm eff}_{2}|\tilde{\gamma}_{{\bf{k}}}|^{2}.
\label{Equation_11}
\end{eqnarray}
The coefficients $\beta_{i{{\bf{k}}}}$ in Eq.(\ref{Equation_9}), with $i=1,..4$ are given by the relations
\begin{eqnarray}
\footnotesize
\arraycolsep=0pt
\medmuskip = 0mu
\beta_{i{{\bf{k}}}}
=(-1)^{i+1}
\left\{
\begin{array}{cc}
\displaystyle  & \frac{\left(\mu^{\rm eff}_{1}-\varepsilon_{i{\bf{k}}}\right)^{2}}{\left(\varepsilon_{1{\bf{k}}}-\varepsilon_{2{\bf{k}}}\right)}\prod^{}_{j=3,4}\frac{1}{\left(\varepsilon_{i{\bf{k}}}-\varepsilon_{j{\bf{k}}}\right)},  \ \ \  $if$ \ \ \ i=1,2,
\newline\\
\newline\\
\displaystyle  & \frac{\left(\mu^{\rm eff}_{1}-\varepsilon_{i{\bf{k}}}\right)^{2}}{\left(\varepsilon_{3{\bf{k}}}-\varepsilon_{4{\bf{k}}}\right)}\prod^{}_{j=1,2}\frac{1}{\left(\varepsilon_{i{\bf{k}}}-\varepsilon_{j{\bf{k}}}\right)},  \ \ \  $if$ \ \ \ i=3,4.
\end{array}\right.
\nonumber\\
\label{Equation_12}
\end{eqnarray}
The function $n_{F}\left(x\right)$, in Eqs.(\ref{Equation_8}) and (\ref{Equation_9}), is the Fermi-Dirac distribution function $n_{F}\left(x\right)=1/\left(e^{\beta(x-\mu)}+1\right)$ and the energy parameters $\varepsilon_{i{\bf{k}}}$ define the excitonic band structure in the interacting BLG. They are given as
\begin{eqnarray}
\varepsilon_{1,2{\bf{k}}}=-\frac{1}{2}\left[\Delta+\gamma_{1}\pm\sqrt{\left(W-\Delta-\gamma_{1}\right)^{2}+4|\tilde{\gamma}_{{\bf{k}}}|^{2}}\right]+\bar{\mu},\
\nonumber\\
\varepsilon_{3,4{\bf{k}}}=-\frac{1}{2}\left[-\Delta-\gamma_{1}\pm\sqrt{\left(W+\Delta+\gamma_{1}\right)^{2}+4|\tilde{\gamma}_{{\bf{k}}}|^{2}}\right]+\bar{\mu}.
\nonumber\\
\label{Equation_13}
\end{eqnarray}
We have introduced the effective Fermi level $\bar{\mu}$ in Eq.(\ref{Equation_13}) in terms of the effective chemical potentials $\mu_{1\rm eff}$ and $\mu_{2\rm eff}$, i.e., $\bar{\mu}=\left(\mu_{1\rm eff}+\mu_{1\rm eff}\right)/2$. The solution of the system of equations given in Eqs.(\ref{Equation_8}) and (\ref{Equation_9}) and the following discussion about the behavior of the excitonic order parameter and chemical potential are well discussed in Refs.\cite{cite_38, cite_41}. In the present work, we will use the data for $\Delta$ and $\mu$, obtained in Ref.\cite{cite_38}, in order to calculate the thermal properties in the suspended bilayer graphene. Just for the completeness, and the logical continuation, we will follow the results given in the work in Ref.\cite{cite_41} and we rewrite here the explicit expressions of the fermionic Green's functions in the interacting bilayer graphene system, under consideration. For the sublattices $A$ and $B$ in the bottom layer in BLG we have for the normal single-particle Green functions \cite{cite_42} ${\cal{G}}_{\eta_{\ell}{\bf{k}}}(\nu_n)=-\left\langle \eta_{\ell{\bf{k}}}(\nu_n)\eta^{\dag}_{\ell{\bf{k}}}(\nu_n)\right\rangle$
\begin{eqnarray}
{\cal{G}}_{a_1{\bf{k}}}(\nu_{n})=-\sum^{4}_{i=1}\frac{\alpha_{i{\bf{k}}}}{i\nu_n+\varepsilon_{i{\bf{k}}}},
\nonumber\\
{\cal{G}}_{b_1{\bf{k}}}(\nu_{n})=-\sum^{4}_{i=1}\frac{\gamma_{i{\bf{k}}}}{i\nu_n+\varepsilon_{i{\bf{k}}}},
\label{Equation_14}
\end{eqnarray}  
where the ${\bf{k}}$-dependent coefficients $\gamma_{i{\bf{k}}}$ are defined as
\begin{eqnarray}
\footnotesize
\arraycolsep=0pt
\medmuskip = 0mu
\gamma_{i{{\bf{k}}}}
=(-1)^{i+1}
\left\{
\begin{array}{cc}
\displaystyle  & \frac{{\cal{P}}'^{(3)}(\varepsilon_{i{\bf{k}}})}{\left(\varepsilon_{1{\bf{k}}}-\varepsilon_{2{\bf{k}}}\right)}\prod^{}_{j=3,4}\frac{1}{\left(\varepsilon_{i{\bf{k}}}-\varepsilon_{j{\bf{k}}}\right)},  \ \ \  $if$ \ \ \ i=1,2,
\newline\\
\newline\\
\displaystyle  & \frac{{\cal{P}}'^{(3)}(\varepsilon_{i{\bf{k}}})}{\left(\varepsilon_{3{\bf{k}}}-\varepsilon_{4{\bf{k}}}\right)}\prod^{}_{j=1,2}\frac{1}{\left(\varepsilon_{i{\bf{k}}}-\varepsilon_{j{\bf{k}}}\right)},  \ \ \  $if$ \ \ \ i=3,4,
\end{array}\right.
\nonumber\\
\label{Equation_15}
\end{eqnarray}
where ${\cal{P}}'^{(3)}(x)$ is a polynomial of third order ${\cal{P}}'^{(3)}(x)=x^{3}+a'_{1{\bf{k}}}x^{2}+a'_{2{\bf{k}}}x+a'_{3\bf{k}}$ with the coefficients $a'_{i{\bf{k}}}$, $i=1,...3$, defined as 
\begin{eqnarray}
&&a'_{1{\bf{k}}}=-2\mu^{\rm eff}_{1}-\mu^{\rm eff}_{2},
   	\nonumber\\
&&a'_{2{\bf{k}}}=\mu^{\rm eff}_{1}\left(\mu^{\rm eff}_{1}+2\mu^{\rm eff}_{2}\right)-|\tilde{\gamma}_{{\bf{k}}}|^{2},
   \nonumber\\
&&a'_{3{\bf{k}}}=-\mu^{\rm eff}_{2}\left(\mu^{\rm eff}_{1}\right)^{2}+\mu^{\rm eff}_{1}|\tilde{\gamma}_{{\bf{k}}}|^{2}.
\label{Equation_16}
\end{eqnarray} 
We see that $a'_{1{\bf{k}}}$, $a'_{2{\bf{k}}}$ and $a'_{3{\bf{k}}}$ could be obtained from the coefficients $a_{1{\bf{k}}}$, $a_{2{\bf{k}}}$ and $a_{3{\bf{k}}}$, given in Eq.(\ref{Equation_11}), by interchanging the effective chemical potentials $\mu^{(1)}_{\rm eff}\rightleftharpoons \mu^{(2)}_{\rm eff}$ and by setting simultaneously $\Delta+\gamma_1=0$. The anomalous or the excitonic Green function in bilayer graphene is defined as ${\cal{F}}_{a_{2}b_{1}{\bf{k}}}(\nu_n)=-\left\langle a_{2{\bf{k}}}(\nu_n)b^{\dag}_{1{\bf{k}}}(\nu_n)\right\rangle$ and we obtain
\begin{eqnarray}
{\cal{F}}_{a_{2}b_{1}{\bf{k}}}(\nu_n)=-\sum^{4}_{i=1}\frac{\beta_{i{\bf{k}}}}{i\nu_n+\varepsilon_{i{\bf{k}}}},
\label{Equation_17}
\end{eqnarray} 
where the coefficients $\beta_{i{\bf{k}}}$ are given in Eq.(\ref{Equation_12}) above. For the layer $\ell=2$, it is not difficult to show that ${\cal{G}}_{a_2{\bf{k}}}(\nu_{n})={\cal{G}}_{b_1{\bf{k}}}(\nu_{n})$ and ${\cal{G}}_{b_2{\bf{k}}}(\nu_{n})={\cal{G}}_{a_1{\bf{k}}}(\nu_{n})$. Furthermore, the Green functions in Eqs.(\ref{Equation_14}) and (\ref{Equation_17}) will be used for the calculation of the thermal kinetic coefficients in bilayer graphene. 
%
\section{\label{sec:Section_3} The electronic and heat current density operators}
%
\begin{figure}
	\begin{center}
		\includegraphics[scale=0.35]{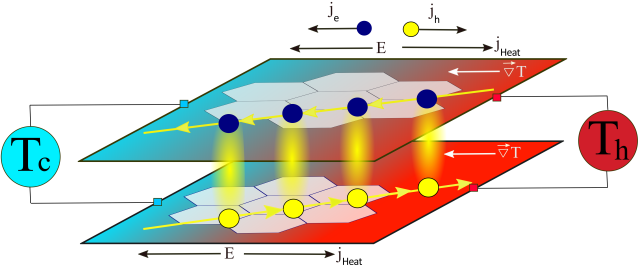}
		\caption{\label{fig:Fig_2}(Color online) The AB-stacked bilayer graphene structure under the temperature gradient. The electric and hole current flow is presented through the stable excitonic condensate state. The directions of the electric field and heat current flow are shown in the picture.}
	\end{center}
\end{figure} 
%
The thermal conductivity is principal for the study of thermal properties in solid state materials. In semiconducting materials with the band gap in the excitation spectrum, the electrons can have a predominant contribution to the heat conduction. It has been shown recently \cite{cite_43} that the thermal transport properties in bilayer graphene could be manipulated by the $sp^{3}$ interlayer bonding and the role of the interlayer phonon coupling has a small impact on the in-plane thermal conductivity in the graphene structures. Here, we consider the role of the excitons on the thermal transport properties in the suspended AB-BLG. We suppose that the temperature gradient along $x$-axis causes an electric field along that axis. In Fig.~\ref{fig:Fig_2}, we have presented the electron and hole currents through the stable excitonic condensate state in the AB-BLG structure. The directions of the temperature gradient $\nabla{T}$, thermoelectric field ${\bf{E}}$, total heat current ${\bf{j}}_{\rm H}$ and electric (${\bf{j}}_{\rm e}$) and hole (${\bf{j}}_{\rm h}$) currents are shown in the picture. The AB-BLG structure is maintained at the constant temperature gradient, between the hot (in red) and cold (in blue) sides of the structure. The charge and heat currents are related within the linear response equations since the temperature gradient $\nabla T$ leads to the electric potential gradient $\nabla V$ in the system. We write here the linear response equations for the electric current density (${{\bf{j}}_{\rm e}}$ ) and total heat current density (${\bf{j}}_{\rm H}$). Then the total heat current density is related to the energy current density ${\bf{j}}_{\rm E}$ via the relation  
\begin{eqnarray}
{\bf{j}}_{\rm H}={\bf{j}}_{\rm E}-\mu{{\bf{j}}_{\rm e}},
\label{Equation_18}
\end{eqnarray}
where $\mu$ is the chemical potential in the system. 
With the definition of thermal current density in Eq.(\ref{Equation_18}), we have a set of coupled linear equations for the electric and heat current density. After defining the appropriate forces for the generation of currents we have in general
\begin{eqnarray}
{\bf{j}}_{\rm e}(\omega)=-\left[\frac{1}{T}(\nabla{\mu})+\frac{{\bf{F}}_{\rm e}(\omega)}{T}\right]{\cal{L}}_{\rm 11}(\omega)-\frac{{\cal{L}}_{\rm 12}(\omega)}{T^{2}}\nabla{T},
\nonumber\\
\label{Equation_19}
\newline\\
{\bf{j}}_{\rm H}(\omega)=-\left[\frac{1}{T}({\bf{\nabla}}{\mu})+\frac{{\bf{F}}_{\rm e}(\omega)}{T}\right]{\cal{L}}_{\rm 21}(\omega)-\frac{{\cal{L}}_{\rm 22}(\omega)}{T^{2}}\nabla{T}.
\nonumber\\
\label{Equation_20}
\end{eqnarray} 
Here, $\nabla{\mu}$ is the gradient of the chemical potential in the system, which is the consequence of the electron concentration gradient. Next, ${\bf{F}}_{\rm e}(\omega)$ is the force acting under effective electric field ${\bf{E}}(\omega)$ produced by the temperature gradient $\nabla{T}$. Thus, ${\bf{F}}_{\rm e}(\omega)=-e{\bf{E}}(\omega)$. The frequency dependent transport coefficients ${\cal{L}}_{\rm ij}(\omega)$ with $i,j =1,2$ are the imaginary parts of retarded response functions \cite{cite_44} which will be calculated later on. With these definitions, the Onsager relation between non-diagonal transport coefficients is ${\cal{L}}_{\rm 12}(\omega)={\cal{L}}_{\rm 21}(\omega)$. Furthermore, we assume that there are no concentration gradients in the considered problem and we put $\nabla{\mu}=0$. We can obtain the analytical expressions of important thermal parameters in the system just by reverting the linear response equations in Eqs.(\ref{Equation_19}) and (\ref{Equation_20}) by writing them in the following equivalent form
\begin{eqnarray}
{\bf{E}}(\omega)=\sigma^{-1}_{\rm e}(\omega){\bf{j}}_{\rm e}(\omega)-S(T)\nabla{T},
\label{Equation_21}
\newline\\
{\bf{j}}_{\rm H}(\omega)=-\Pi{\bf{j}}_{\rm e}(\omega)-\kappa\nabla{T},
\label{Equation_22}
\end{eqnarray} 
where $\sigma_{\rm e}(\omega)$ is the electrical conductivity, caused by the temperature gradient, $S(T)$ is the Seebeck coefficient, $\Pi$ is the Peltier coefficient and $\kappa$ is the thermal conductivity which is defined as the coefficient of proportionality (when the electric current density is zero) between the total heat current density and the temperature gradient, i.e.,  ${\bf{j}}_{\rm H}(\omega)=-\kappa\nabla{T}$. We have
\begin{eqnarray}    
&&\sigma_{\rm e}(\omega)=\frac{{\cal{L}}_{\rm 11}(\omega)}{T},
\nonumber\\
&&S(T)=-\frac{{\cal{L}}_{\rm 12}(\omega)}{T{\cal{L}}_{\rm 11}(\omega)},
\nonumber\\
&&\Pi(T)=T\cdot{S(T)},
\nonumber\\
&&\kappa(T)=\frac{1}{T^{2}}\left({\cal{L}}_{\rm 22}(\omega)-\frac{{\cal{L}}^{2}_{\rm 12}(\omega)}{{\cal{L}}_{\rm 11}(\omega)}\right).
\label{Equation_23}
\end{eqnarray} 
In the definitions, above, we have putted $e\equiv 1$. The thermoelectric conversion efficiency is evaluated with the help of the thermoelectric figure of merit $ZT$ which is the regrouped combination of the principal thermoelectric parameters $S(T), \sigma_{\rm e}$ and $\kappa$
\begin{eqnarray}    
ZT=S^{2}\sigma{T}/\kappa.  
\label{Equation_24}
\end{eqnarray} 
The thermal conductivity $\kappa$ is given in general as $\kappa=\kappa_{\rm el}+\kappa_{\rm ph}$ with the electronic ($\kappa_{\rm el}$) and phononic ($\kappa_{\rm ph}$) contributions.
The thermoelectric efficiency represents the ratio between the thermal power input and the electrical power output \cite{cite_45}, for a given material, and the maximum value of it is defined as \cite{cite_46}
\begin{eqnarray}    
\eta_{\rm eff}=\frac{\Delta{T}}{T_{\rm h}}\frac{\sqrt{1+{{ZT}}_{\rm av}}-1}{\sqrt{1+{{ZT}}_{\rm av}}+\frac{T_{\rm c}}{T_{\rm h}}},
\label{Equation_25}
\end{eqnarray}
where $T_{\rm h}$ and $T_{\rm c}$ are the temperatures in the hot and cold sides of the material and the temperature difference is $\Delta{T}=T_{\rm h}-T_{\rm c}$. The average $T_{\rm av}$ is the mean operating temperature in the sample and we have ${T}_{\rm av}=(T_{\rm h}+T_{\rm c})/2$. Thus the thermoelectric conversion efficiency is the product of the Carnot efficiency $\eta_{c}=(\Delta{T}/T_{h})$ and a reduction factor which is the function of the material's figure of merit parameter, defined as $Z=S^{2}\sigma/\kappa$. It is important to notice here that $50 \%$ of the waste heat and the solar thermal energy lie in the range $300$-$500$ K, while no efficient thermoelectric materials have been found for low-temperature applications ($<500$ K). For example, a thermoelectric generator, composed of a material with the FOM $ZT =1$, is expected to reach only $5\%$ energy conversion efficiency when $\Delta{T}=100$ K. Most research efforts are focused on materials
development with the increasing FOM $ZT$ and the maximum $\Delta{T}$ which is highlighted by the Carnot efficiency $\eta_{\rm c}$. Recently, it has been shown that the thermoelectric power in monolayer and bilayer graphene follows the semiclassical Mott formula \cite{cite_15, cite_47} at the low-temperature limit, while at the high carrier density the temperature dependence was shown to be linear, which demonstrates the absence of phonon drag modes in the bilayer graphene structure. Regarding this, we will neglect the phononic part of thermal conductivity $\kappa_{\rm ph}$ and further derivations concern only to the electrons. As it has been mentioned earlier, we consider the longitudinal thermal transport in bilayer graphene by supposing the temperature gradient along $x$-direction in the planes of the BLG. 
In order to evaluate the thermal parameters given in Eq.(\ref{Equation_23}) we should calculate the linear response coefficients ${\cal{L}}_{\rm ij}(\omega)$ $i,j=1,2$ which are diagonal in $i,j$ for the isotropic systems. They are given with the help of the Kubo formula, and, in the Matsubara notations, we have for ${\cal{L}}_{\rm ij}(\omega_{m})$, ($\omega_{m}$ here are the bosonic Matsubara frequencies $\omega_{m}=2\pi{m}/\beta$) along $x$-direction 
\begin{eqnarray}    
{\cal{L}}_{\rm ij}(\omega_{m})=\frac{1}{\beta{\omega_{m}}}\int^{\beta}_{0}d\tau{e^{i\omega_{m}\tau}}\left\langle T_{\tau}{j}_{\rm ix}(\tau){j}_{\rm jx}(0)\right\rangle.
\label{Equation_26}
\end{eqnarray}
The dynamical transport coefficients in Eqs.(\ref{Equation_23}), (\ref{Equation_24}) and Eq.(\ref{Equation_26}) are obtained after the analytical continuation in Eq.(\ref{Equation_26}) $\i\omega_{m}\rightarrow \omega+i\delta$, where $\omega$ are the real frequencies and $\delta$ is an infinitesimal constant \cite{cite_44}. We have
\begin{eqnarray}    
{\cal{L}}_{\rm ij}(\omega)=\Im{{\cal{L}}_{\rm ij}(i\omega_{m}\rightarrow \omega+i\delta)}.
\label{Equation_27}
\end{eqnarray}
Here, in this paper, we will omit the notation of the components along $x$ for the principal thermoelectric parameters in Eq.(\ref{Equation_28}) and the response coefficients ${\cal{L}}_{\rm ij}(\omega)$.
After Eqs.(\ref{Equation_19}) and (\ref{Equation_20}), the current density operators in the time-ordered correlation function in Eq.(\ref{Equation_26}) are defined by the convention that ${j}_{\rm 1x}={j}_{\rm ex}$ and ${j}_{\rm 2x}={j}_{\rm Hx}$, where ${j}_{\rm ex}$ and ${j}_{\rm Hx}$ are the electronic and heat current density operators. To calculate the current-current correlation functions in Eq.(\ref{Equation_26}), for a given response function, we should know the explicit expressions of the electronic and heat current density operators in our bilayer graphene system. The calculation of those operators could be done within the Mahan's formalism evaluated in Ref.\cite{cite_48}, for the electric and heat current density operators.  
Thus, for the electric current density operator, we introduce the electron polarization operator ${\bf{P}}_{\rm e}$, which is the sum over the position ${\bf{R}}_{{\bf{m}}}$ (of a lattice site ${\bf{m}}$) and the electron number operator $n_{{\bf{m}}}$ at that site, i.e., ${\bf{P}}_{\rm e}=\sum_{{\bf{m}}}{\bf{R}}_{{\bf{m}}}n_{{\bf{m}}}$, while for the energy current density operator ${j}_{\rm E}$ we construct an operator ${\bf{R}}_{E}$ which is the sum over position ${\bf{R}}_{{\bf{m}}}$ and effective Hamiltonian density $h_{\rm eff{\bf{m}}}$ (see in \ref{sec:Section_7}), i.e., ${\bf{R}}_{E}=\sum_{{\bf{m}}}{\bf{R}}_{{\bf{m}}}h_{\rm eff{\bf{m}}}$. Furthermore, by using the continuity equations and the Heisenberg equation of motion for the current density operators it is easy to show that the electron and heat current density operators satisfy the following equations 
\begin{eqnarray}    
&&{j}_{\rm ex}=\frac{\partial {P}_{\rm x}}{\partial t}=i\sum_{{\bf{m}},{\bf{n}}}{\bf{R}}_{\rm {\bf{m}}x}\left[h_{\rm eff{\bf{n}}},n_{{\bf{m}}}\right],                     
\newline\\
&&{j}_{\rm Ex}=\frac{\partial {R}_{\rm Ex}}{\partial t}=i\sum_{{\bf{m}},{\bf{n}}}{\bf{R}}_{\rm {\bf{m}}x}\left[h_{\rm eff{\bf{n}}},h_{\rm eff{\bf{m}}}\right].
\label{Equation_28} 
\end{eqnarray}
Here, ${\bf{R}}_{\rm {\bf{m}}x}$ means the $x$-component of the position vector ${\bf{R}}_{{\bf{m}}}$ at the lattice site ${\bf{m}}$. Furthermore, after some calculations (see in \ref{sec:Section_7}) we can write the electronic and heat current density operators in the following forms
\begin{eqnarray}    
{j}_{\rm ex}=\frac{1}{N}\sum_{{\bf{k}}\nu_n,\sigma}\Psi^{\dag}_{{\bf{k}}\sigma}(\nu_{n}){\upsilon}_{{\bf{k}}x}\Psi_{{\bf{k}}\sigma}(\nu_{n}),
\label{Equation_29} 
\end{eqnarray}
where the fermionic spinors $\Psi^{\dag}_{{\bf{k}}\sigma}(\nu_{n})$ and $\Psi_{{\bf{k}}\sigma}(\nu_{n})$ have been introduced in Eq.(\ref{Equation_5}), in the Section \ref{sec:Section_2_1}, while the electron velocity matrix operator ${\upsilon}_{{\bf{k}}x}$ is defined as (see in \ref{sec:Section_7}, for details)
\begin{eqnarray}
{\upsilon}_{{\bf{k}}x}=\left(
\begin{array}{ccccrrrr}
0 & 0 & 0 & {\bf{v}}_{{\bf{k}}x}\\
0 &0  & {\bf{v}}_{{\bf{k}}x} & 0 \\
0 & {\bf{v}}^{\ast}_{{\bf{k}}x} & 0 & 0 \\
{\bf{v}}^{\ast}_{{\bf{k}}x} & 0 & 0 & 0
\end{array}
\right).
\label{Equation_30}
\end{eqnarray}
Here ${\bf{v}}_{{\bf{k}}x}$ and ${\bf{v}}^{\ast}_{{\bf{k}}x}$ are the $x$-components of the ${\bf{k}}$-dependent electron velocity operator ${\bf{v}}_{{\bf{k}}}$ and its complex conjugate. We have defined here
\begin{eqnarray}    
{\bf{v}}_{\bf{k}}=-\frac{i\gamma_0}{\hbar}\sum_{{\bm{\mathit{\delta}}}_{1}}{\bm{\mathit{\delta}}}_{1}e^{i{\bf{k}}{\bm{\mathit{\delta}}}_{1}}.
\label{Equation_31} 
\end{eqnarray}
The total heat current density operator along the $x$ direction could be calculated as ${j}_{\rm Hx}={j}_{\rm Ex}-\mu{j}_{\rm ex}$. After calculating the commutator $\left[H,{R}_{\rm E}\right]$ (see in \ref{sec:Section_7_2}, for details), we get
\begin{eqnarray}    
{j}_{\rm Hx}=\frac{1}{N}\sum_{{\bf{k}}\nu_{n},\sigma}\Psi^{\dag}_{{\bf{k}}\sigma}(\nu_{n}){\bf{\upsilon}}^{\rm H}_{{\bf{k}}x}\Psi_{{\bf{k}}\sigma}(\nu_{n}),
\label{Equation_32} 
\end{eqnarray}
where the total velocity matrix operator ${\bf{\upsilon}}^{\rm H}_{{\bf{k}}x}$ is defined as
\begin{eqnarray}
{\bf{\upsilon}}^{\rm H}_{{\bf{k}}x}=\left(
\begin{array}{ccrr}
{\bf{\upsilon}}_{g{\bf{k}}} & {\bf{\upsilon}}_{t{\bf{k}}} \\
{\bf{\upsilon}}^{\dag}_{t{\bf{k}}} & {\bf{\upsilon}}_{g{\bf{k}}}   \\
\end{array}
\right)_{\rm x}.
\label{Equation_33}
\end{eqnarray}
Each element of the matrix ${\bf{\upsilon}}^{\rm H}_{{\bf{k}}x}$ in Eq.(\ref{Equation_33}) represents a $2\times 2$ matrix, and we have 
\begin{eqnarray}
{\bf{\upsilon}}_{g{\bf{k}}}=\left(
\begin{array}{ccrr}
{\bf{v}}_{1{\bf{k}}} & 0 \\
0 & 0   \\
\end{array}
\right),
\label{Equation_34}
\end{eqnarray}  
\begin{eqnarray}
{\bf{\upsilon}}_{t{\bf{k}}}=\left(
\begin{array}{ccrr}
{\bf{v}}_{2{\bf{k}}} & {\bf{v}}_{4{\bf{k}}} \\
{\bf{v}}_{3{\bf{k}}} & {\bf{v}}_{2{\bf{k}}}   \\
\end{array}
\right).
\label{Equation_35}
\end{eqnarray} 
The matrix ${\bf{\upsilon}}^{\dag}_{t{\bf{k}}}$ in Eq.(\ref{Equation_33}) is the Hermitian conjugate of the matrix ${\bf{\upsilon}}_{t{\bf{k}}}$. The elements of the matrices ${\bf{\upsilon}}_{g{\bf{k}}}$ and ${\bf{\upsilon}}_{t{\bf{k}}}$ are introduced in the following form 
\begin{eqnarray}    
&&{\bf{v}}_{1{\bf{k}}}=\gamma_0\Lambda_{{\bf{k}}},
\nonumber\\
&&{\bf{v}}_{2{\bf{k}}}=-(\gamma_1+\Delta){\bf{v}}_{{\bf{k}}},
\nonumber\\
&&{\bf{v}}_{3{\bf{k}}}=\left(\mu_{\rm 1eff}-\mu\right){\bf{v}}_{{\bf{k}}},
\nonumber\\
&&{\bf{v}}_{4{\bf{k}}}=-\left(\mu_{\rm 2eff}+\mu\right){\bf{v}}_{{\bf{k}}}.
\label{Equation_36} 
\end{eqnarray}
The function $\Lambda_{{\bf{k}}}$ in the expression of ${\bf{v}}_{1{\bf{k}}}$ is
\begin{eqnarray}    
\Lambda_{{\bf{k}}}=-i\gamma_0\sum_{{\bf{u}}}{\bf{u}}e^{i{\bf{k}}{\bf{u}}},
\label{Equation_37} 
\end{eqnarray}
where the summation vector ${\bf{u}}$ takes the values $0, \pm {\bf{a}}_{1}, \pm {\bf{a}}_{2}$ with the vectors ${\bf{a}}_{1}$ and ${\bf{a}}_{2}$ being the unit cell vectors in graphene lattice structure
\begin{eqnarray}    
{\bf{a}}_{1}=\bm{\mathit{\delta}}^{(1)}_{1}-\bm{\mathit{\delta}}^{(3)}_{1}=\frac{3a}{2}\left( 1,\frac{1}{\sqrt{3}}\right),
\newline\\
{\bf{a}}_{2}=\bm{\mathit{\delta}}^{(2)}_{1}-\bm{\mathit{\delta}}^{(3)}_{1}=\frac{3a}{2}\left( 1,\frac{1}{\sqrt{3}}\right).
\label{Equation_38} 
\end{eqnarray}
%
\section{\label{sec:Section_4} Thermal linear response functions}
%
Here, we give the results of calculations of the response functions using the
definitions in Eqs.(\ref{Equation_26}) and (\ref{Equation_27}). For this, we put the expressions for the electronic and heat current densities in Eqs.(\ref{Equation_29}) and (\ref{Equation_32}) into the expression of the correlation function in Eq.(\ref{Equation_26}) and we apply the Wick's theorem \cite{cite_42} for the statistical average of the product of four fermionic operators. Then, after using the definitions of the normal and excitonic Green's functions in Eqs.(\ref{Equation_14}) and (\ref{Equation_17}), we obtain 
\begin{eqnarray}    
&&{\cal{L}}_{\rm 11}(\omega_{m})=-\frac{2}{N\beta^{2}\omega_{m}}\sum_{{\bf{k}},\nu_{n}}\sum_{\sigma}\left({\cal{G}}^{\sigma}_{a_{1}{\bf{k}}}(\nu_{n}){\cal{G}}^{\sigma}_{b_{1}{\bf{k}}}(\nu_{n}+\omega_{m})\right.
\nonumber\\
&&\left. +{\cal{G}}^{\sigma}_{a_{1}{\bf{k}}}(\nu_{n}){\cal{G}}^{\sigma}_{b_{1}{\bf{k}}}(\nu_{n}-\omega_{m})\right)|{\bf{v}}_{{\bf{k}}x}|^{2}.
\label{Equation_39} 
\end{eqnarray}
Furthermore, we perform the summation over fermionic Matsubara frequencies $\nu_{n}$ in Eq.(\ref{Equation_39}) and the analytical continuation, given in Eq.(\ref{Equation_27}), in order to obtain the expression of the retarded real-frequency response function. After taking the imaginary part of the retarded function, we get
\begin{eqnarray}    
&&{\cal{L}}_{\rm 11}(\omega)=\frac{4\pi}{N\beta\omega}\sum_{{\bf{k}}}\sum^{4}_{\rm i,j=1}\alpha_{i{\bf{k}}}\gamma_{j{\bf{k}}}\left(n_{\rm F}(-\epsilon_{i{\bf{k}}})-n_{\rm F}(-\epsilon_{j{\bf{k}}})\right)
\nonumber\\
&&\times \left(\delta(\epsilon_{j{\bf{k}}}-\epsilon_{i{\bf{k}}}+\omega)-\delta(\epsilon_{j{\bf{k}}}-\epsilon_{i{\bf{k}}}-\omega)\right)|{\bf{v}}_{{\bf{k}}x}|^{2}.
\label{Equation_40} 
\end{eqnarray}  
For the non-diagonal element ${\cal{L}}_{\rm 12}(\omega_{m})$ we have
\begin{eqnarray}    
&&{\cal{L}}_{\rm 12}(i\omega_{m})=\frac{1}{N\beta^{2}\omega_{m}}\sum_{{\bf{k}},\nu_{n}}\sum_{\sigma}\left[2{\cal{G}}^{\sigma}_{a_{1}{\bf{k}}}(\nu_{n}){\cal{F}}^{\sigma}_{a_{2}b_{1}{\bf{k}}}(\nu_{n}+\omega_{m})\right.
\nonumber\\
&&\left.\times(\gamma_1+\Delta)+{\cal{G}}^{\sigma}_{a_{1}{\bf{k}}}(\nu_{n}){\cal{G}}^{\sigma}_{b_{1}{\bf{k}}}(\nu_{n}+\omega_{m})(W+2\mu)+\right.
\nonumber\\
&&\left.+ 2{\cal{G}}^{\sigma}_{a_{1}{\bf{k}}}(\nu_{n}){\cal{F}}^{\sigma}_{a_{2}b_{1}{\bf{k}}}(\nu_{n}-\omega_{m})(\gamma_1+\Delta)+\right.
\nonumber\\
&&\left. +{\cal{G}}^{\sigma}_{a_{1}{\bf{k}}}(\nu_{n}){\cal{G}}^{\sigma}_{b_{1}{\bf{k}}}(\nu_{n}-\omega_{m})(W+2\mu)
\right]|{\bf{v}}_{{\bf{k}}x}|^{2}.
\label{Equation_41} 
\end{eqnarray} 
Again, after performing the Matsubara summation in Eq.(\ref{Equation_41}) we get for the real-frequency dependent coefficient ${\cal{L}}_{\rm 12}(\omega)$
\begin{eqnarray}    
&&{\cal{L}}_{\rm 12}(\omega)=-\frac{2\pi}{N\beta\omega}\sum_{{\bf{k}}}\sum^{4}_{\rm i,j=1}\lbrace\left[2\alpha_{i{\bf{k}}}\beta_{j{\bf{k}}}(\gamma_1+\Delta)^{2}\right.
\nonumber\\
&&\left. +\alpha_{i{\bf{k}}}\gamma_{j{\bf{k}}}(W+2\mu)\right]\left(n_{\rm F}(-\epsilon_{i{\bf{k}}})-n_{\rm F}(-\epsilon_{j{\bf{k}}})\right)
\nonumber\\
&& \times \left(\delta(\epsilon_{j{\bf{k}}}-\epsilon_{i{\bf{k}}}+\omega)-\delta(\epsilon_{j{\bf{k}}}-\epsilon_{i{\bf{k}}}-\omega)\right)\rbrace
|{\bf{v}}_{{\bf{k}}x}|^{2}.
\label{Equation_42} 
\end{eqnarray} 
The coefficient ${\cal{L}}_{\rm 21}(\omega)$ is the same as ${\cal{L}}_{\rm 12}(\omega)$. The second diagonal coefficient ${\cal{L}}_{\rm 22}(\omega)$ could be also calculated using Eq.(\ref{Equation_32}) for the total heat current density operator. The analytical expression of the coefficient ${\cal{L}}_{\rm 22}(\omega_{m})$, in terms of the normal and excitonic Green functions is given in \ref{sec:Section_8}. Here, we present only the real frequency value of it, after analytical continuation into upper half complex plane $i\omega_{m}\rightarrow \omega+i\delta$  
\begin{eqnarray}    
&&{\cal{L}}_{\rm 22}(\omega)=\frac{2\pi}{N\beta\omega_{m}}\sum_{{\bf{k}}}\sum^{4}_{\rm i,j=1}\left[
\alpha_{i{\bf{k}}}\alpha_{j{\bf{k}}}+\gamma_{i{\bf{k}}}\gamma_{j{\bf{k}}}\right]\times
\nonumber\\
&&\delta\left(\epsilon_{j{\bf{k}}}-\epsilon_{i{\bf{k}}}+\omega\right)\left(n_{\rm F}(-\epsilon_{i{\bf{k}}})-n_{\rm F}(-\epsilon_{j{\bf{k}}})\right)|{\bf{v}}_{1{\bf{k}}x}|^{2}
\nonumber\\
&&+\left[\left(
\alpha_{i{\bf{k}}}\gamma_{j{\bf{k}}}+\gamma_{i{\bf{k}}}\alpha_{j{\bf{k}}}\right)(\gamma_1+\Delta)^{2}+\right.
\nonumber\\
&&\left.+2\alpha_{i{\bf{k}}}\beta_{j{\bf{k}}}(\gamma_1+\Delta)(\mu_{\rm 2eff}+\mu)+\right.
\nonumber\\
&&\left.+2\beta_{i{\bf{k}}}\alpha_{j{\bf{k}}}(\gamma_1+\Delta)(\mu-\mu_{\rm 1eff})+\alpha_{i{\bf{k}}}\gamma_{j{\bf{k}}}(\mu_{\rm 2eff}+\mu)^{2}+\right.
\nonumber\\
&&\left.+\gamma_{i{\bf{k}}}\alpha_{j{\bf{k}}}(\mu_{\rm 1eff}-\mu)^{2}\right]\left(n_{\rm F}(-\epsilon_{i{\bf{k}}})-n_{\rm F}(-\epsilon_{j{\bf{k}}})\right)
\nonumber\\
&&\times\left(\delta(\epsilon_{j{\bf{k}}}-\epsilon_{i{\bf{k}}}+\omega)-\delta(\epsilon_{j{\bf{k}}}-\epsilon_{i{\bf{k}}}-\omega)\right)|{\bf{v}}_{{\bf{k}}x}|^{2}.
\label{Equation_43} 
\end{eqnarray} 
The $x$-component of the function ${\bf{v}}_{1{\bf{k}}}$ in the first term in Eq.(\ref{Equation_43}), could be evaluated easily. Taking into account the definition in Eq.(\ref{Equation_36}), we have 
\begin{eqnarray}    
&&{\bf{v}}_{1{\bf{k}}x}=3a\gamma^{2}_{0}\left(\sin\left(\frac{3k_{x}{a}}{2}+\frac{\sqrt{3}k_{y}{a}}{2}\right)+\right.
\nonumber\\
&&\left.+\sin\left(\frac{3k_{x}{a}}{2}-\frac{\sqrt{3}k_{y}{a}}{2}\right)\right).
\label{Equation_44} 
\end{eqnarray}
Thus, for the ${\bf{k}}$-dependent coefficient $|{\bf{v}}_{1{\bf{k}}x}|^{2}$, in the first term in Eq.(\ref{Equation_43}), we obtain
\begin{eqnarray}    
|{\bf{v}}_{1{\bf{k}}x}|^{2}=36a^{2}\gamma^{4}_{0}\sin^{2}\frac{3k_{x}{a}}{2}\cos^{2}\frac{\sqrt{3}k_{y}{a}}{2}.
\label{Equation_45} 
\end{eqnarray} 
Next, the $x$ component of the ${\bf{k}}$-dependent velocity operator ${\bf{v}}_{{\bf{k}}}$, defined in Eq.(\ref{Equation_31}), is 
\begin{eqnarray}    
{\bf{v}}_{{\bf{k}}x}=-\frac{i\gamma_0{a}}{\hbar}\left(e^{i\frac{k_{x}a}{2}}\cos{\frac{\sqrt{3}k_{y}{a}}{2}}-e^{-ik_{x}{a}}\right)
\label{Equation_46} 
\end{eqnarray} 
and we have
\begin{eqnarray}    
&&|{\bf{v}}_{{\bf{k}}x}|^{2}=\left(\frac{\gamma_{0}a}{\hbar}\right)^{2}\left(\cos^{2}\frac{\sqrt{3}k_{y}{a}}{2}\right.
\nonumber\\
&&\left.-2\cos\frac{3k_{x}{a}}{2}\cos\frac{\sqrt{3}k_{y}a}{2}+1\right).
\label{Equation_47} 
\end{eqnarray} 
Furthermore, when calculating numerically the response coefficients ${\cal{L}}_{\rm ij}(\omega)$, we will replace $\delta$-Dirac functions, figuring in Eqs.(\ref{Equation_40}), (\ref{Equation_42}) and (\ref{Equation_43}), by a $\delta$-Lorentzian with the appropriate broadening parameter.
%
\section{\label{sec:Section_5} The results and discussion}
%
\subsection{\label{sec:Section_5_1} The thermal conductivity}
\begin{figure}
	\begin{center}
		\includegraphics[scale=0.58]{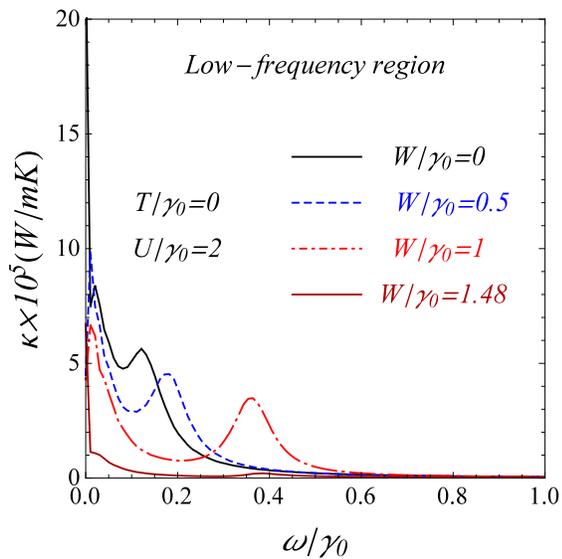}
		\caption{\label{fig:Fig_3}(Color online) Thermal conductivity function, given in Eq.(\ref{Equation_23}) as a function of the incident photon's frequency normalized to the intralayer hopping parameter $\gamma_0$. The zero temperature limit is considered and the low-frequency region is shown. Different values of the interaction parameter $W<W_{\rm CNP}=1.4899\gamma_{0}$ are considered.}
	\end{center}
\end{figure}

%
In this section, we discuss the numerical results concerning the thermal conductivity in the bilayer graphene. Particularly, we will show the interaction induced excitonic effects in the behavior of this function. 
In Fig.~\ref{fig:Fig_3}, we have shown the low-frequency behavior of the thermal conductivity function $\kappa(\omega)$. Different values of the interaction parameter $W$ have been considered and the zero temperature limit is considered. We see in Fig.~\ref{fig:Fig_3} that there are double peaks, appearing at the low-frequency region. When increasing the interlayer Coulomb interaction parameter $W$, the positions of those peaks are displacing to the right (the ``blue"-shift effect in the spectrum) on the $\omega$-axis and also the thermal conductivity is decreasing when augmenting $W$. The low frequency peaks disappear when $W$ is crossing the critical value at the charge neutrality point (CNP) $W_c=1.4899\gamma_0$ (for details see in Refs.\cite{cite_38} and \cite{cite_41}). The similar single peaks, in the low-frequency region, have been observed in Ref.\cite{cite_23} concerning the biased $AA$-stacked bilayer graphene, studied within the tight-binding model Hamiltonian. Particularly, it has been shown that the intensities of peaks decrease when the temperature increased. 
%
%
The doubling of peaks in the case of our AB-stacked bilayer graphene could be related to the excitonic effects considered here. The opposite behavior was shown for $\kappa(\omega)$ when varying the applied bias voltage (see in Ref.\cite{cite_23}). It is important to note that at the value $W=1.48\gamma_0$, close to the CNP value $W_{\rm CNP}=1.4899\gamma_0$, the thermal conductivity is very small (see the dark-red curve in Fig.~\ref{fig:Fig_3}). 
%
\begin{figure}
	\begin{center}
		\includegraphics[scale=0.58]{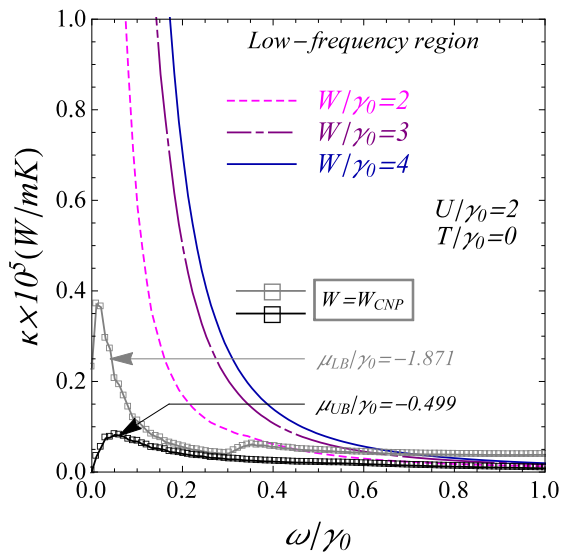}
		\caption{\label{fig:Fig_4}(Color online) Thermal conductivity function, given in Eq.(\ref{Equation_23}) as a function of the incident photon's frequency normalized to the intralyer hopping parameter $\gamma_0$. The zero temperature limit is considered and the low-frequency region is shown. Different values of the interaction parameter  $W\geq W_{\rm CNP}$ are considered in the picture. Lower and upper values of the chemical potential solutions are shown at the CNP point.}
	\end{center}
\end{figure} 
%
At the CNP point, the exact solution for the chemical potential in the AB-BLG, discussed in Ref.\cite{cite_38}, jumps into its upper bound solution branch $\mu_{\rm LB}\rightarrow \mu_{\rm UB}$ (such a behavior of the chemical potential has been observed also experimentally in Ref.\cite{cite_49}), and there are two solutions for $\mu$ and the excitonic order parameter $\Delta$ at $W=W_{\rm CNP}$(this becomes clear when solving the system of coupled equations in Eqs.(\ref{Equation_8}) and (\ref{Equation_9})). One of those solutions is the lower bound solution (LB): $\mu_{\rm LB}=-1.871\gamma_0$ and $\Delta_{\rm LB}=0.0364\gamma_0$, and the another one is the upper bound solution with $\mu_{\rm UB}=-0.499\gamma_0$ and $\Delta_{\rm UB}=0.0249\gamma_0$. 
%
\begin{figure}
	\begin{center}
		\includegraphics[scale=0.58]{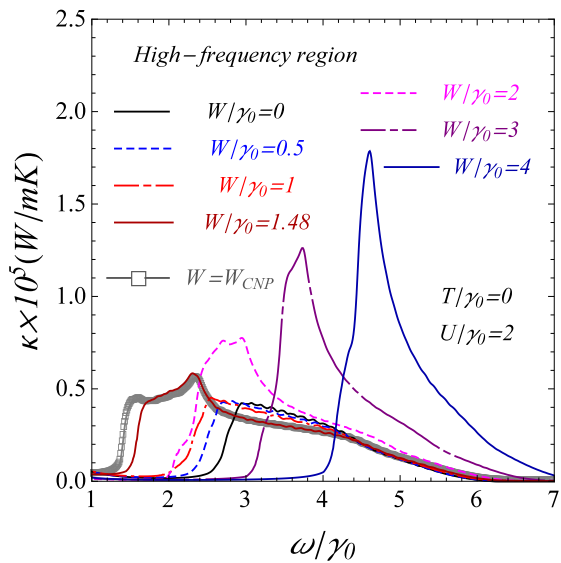}
		\caption{\label{fig:Fig_5}(Color online) Thermal conductivity function, given in Eq.(\ref{Equation_23}) as a function of the incident photon's frequency normalized to the intralyer hopping parameter $\gamma_0$. The zero temperature limit is considered and the high-frequency region is shown. The full bandwidth of the interaction parameter $W$ has been shown in the picture, covering the low and strong interaction limits in the AB-BLG.}
	\end{center}
\end{figure} 
%
Furthermore, at the CNP, $\kappa(\omega)$ becomes very small, especially at the upper bound solution for the chemical potential $\mu_{\rm UB}$, shown in the picture, in Fig.~\ref{fig:Fig_4}. We see in Fig.~\ref{fig:Fig_4} that the double-peak structure in the thermal conductivity function practically disappears at the value $W=W_{\rm CNP}=1.4899\gamma_0$. For the values of $W$ higher than $W_{\rm CNP}$ the double-peak structure is completely absent. 
The spectrum of the electronic thermal conductivity function at the high photon's frequencies is shown in Fig.~\ref{fig:Fig_5}, where we considered the full bandwidth of the interlayer Coulomb interaction parameter, passing through the low and strong interaction limits. We observe in Fig.~\ref{fig:Fig_5} the additional and very broad peaks at the high frequency region. In this case, there is a large``blue"-shift broadening of thermal conductivity when increasing $W$ at $W>W_{\rm CNP}$ while at the values $W<W_{\rm CNP}$ a small ``red"-shift effect is also visible (see the curves corresponding to $W=0$, $W=0.5\gamma_0$, $W=\gamma_0$ and $W=1.48\gamma_0$). At the CNP point, we have shown only the curve corresponding to $\mu_{\rm UB}$, because the other plot for $\mu_{\rm LB}$ is very similar. The existence of thermal conductivity peaks in the high frequency domain could be explained as the result of the high-frequency excitonic effects recently observed in bilayer graphene when considering the optical conductivity in this system \cite{cite_40, cite_50,cite_51}. The small values of $\kappa(\omega)$ imply the large values for the thermoelectric figure of merit, thus, the relatively small values of the thermal conductivity in the  high frequency region, as compared with those in the low frequency regime, could be interesting for the thermoelectrical device applications of the BLG.   
%
\subsection{\label{sec:Section_5_2} The figure of merit and electrical conductivity}
\begin{figure}
	\begin{center}
		\includegraphics[scale=0.58]{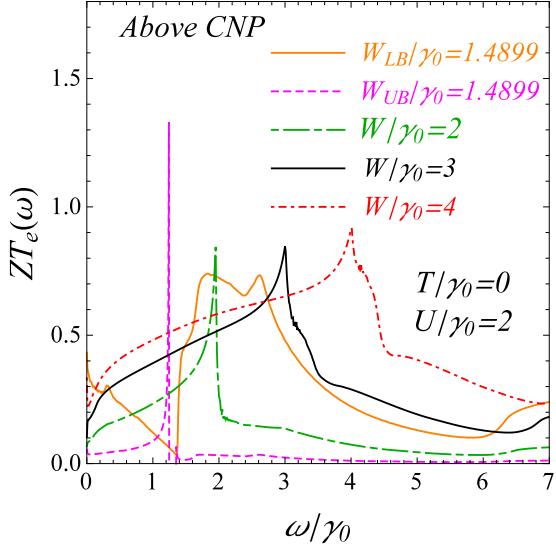}
		\caption{\label{fig:Fig_7}(Color online) The frequency dependence of the electronic thermoelectric figure of merit (FOM), given in Eq.(\ref{Equation_48}). Different values of the interlayer interaction parameter are considered ($W\geq W_{\rm CNP}$).}
	\end{center}
\end{figure}

\begin{figure}
	\begin{center}
		\includegraphics[scale=0.58]{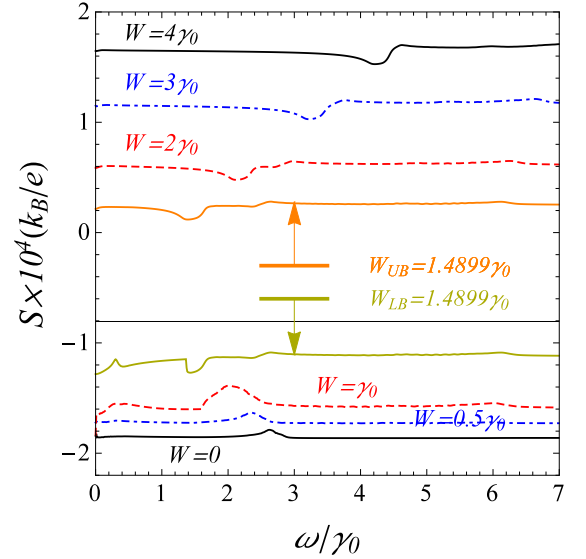}
		\caption{\label{fig:Fig_8}(Color online) The frequency dependence of the Seebeck coefficient in bilayer graphene, given in Eq.(\ref{Equation_23}). Different values of the interlayer interaction parameter are considered, below and above the CNP point.}
	\end{center}
\end{figure}

\begin{figure}
	\begin{center}
		\includegraphics[scale=0.58]{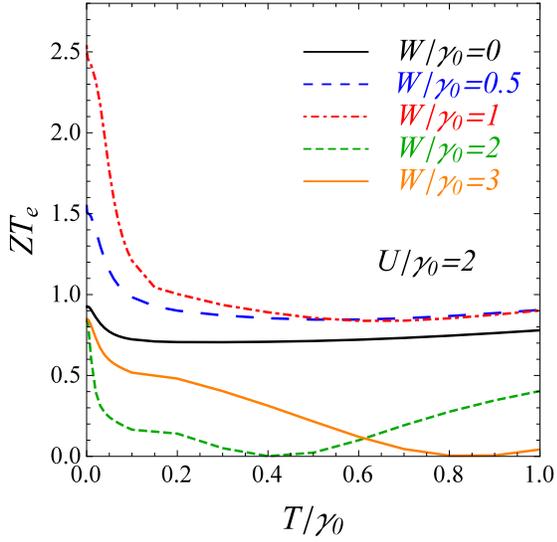}
		\caption{\label{fig:Fig_9}(Color online) The temperature dependence of the thermoelectric figure of merit in bilayer graphene, given in Eq.(\ref{Equation_48}). Different values of the interlayer interaction parameter are considered below and above the CNP point.}
	\end{center}
\end{figure}

\begin{figure}
	\begin{center}
		\includegraphics[scale=0.58]{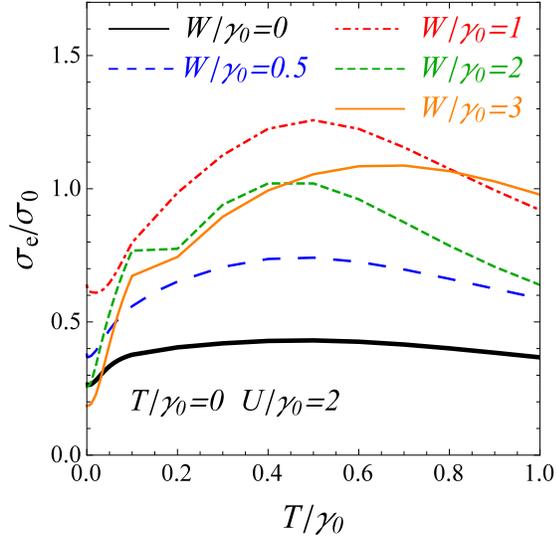}
		\caption{\label{fig:Fig_10}(Color online) The temperature dependence of the electrical conductivity function in bilayer graphene, given in Eq.(\ref{Equation_23}). Different values of the interlayer interaction parameter are considered below and above the CNP point. The conductivity function is normalized at the value $\sigma_{0}=e^{2}/\hbar$.}
	\end{center}
\end{figure}

\begin{figure}
	\begin{center}
		\includegraphics[scale=0.58]{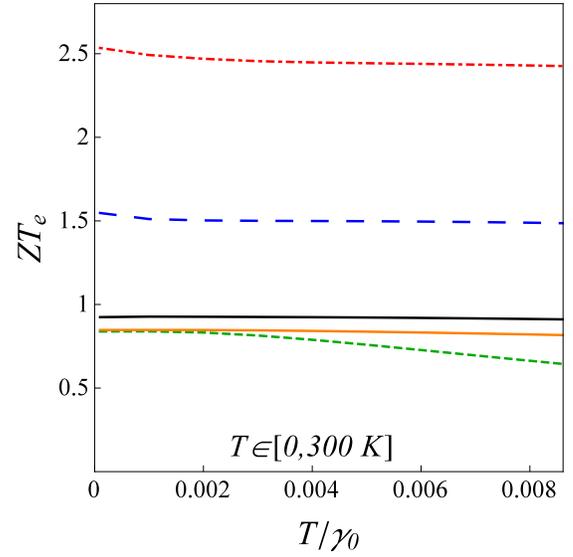}
		\caption{\label{fig:Fig_11}(Color online)  The temperature dependence of the figure of merit in bilayer graphene, given in Eq.(\ref{Equation_48}). The region starting from zero up to room temperature $T=300$ K ($T=0.0258$ eV) is shown in the picture for the same values of the interaction parameter given in Fig.~\ref{fig:Fig_9}.}
	\end{center}
\end{figure}

\begin{figure}
	\begin{center}
		\includegraphics[scale=0.58]{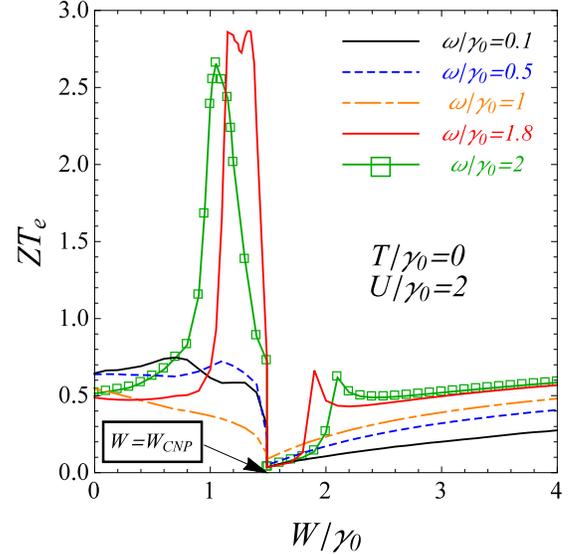}
		\caption{\label{fig:Fig_12}(Color online) The thermoelectric FOM as a function of the interlayer Coulomb interaction parameter, for different values of the incident photon's frequencies. The arrow in black indicates the position of the charge neutrality point on the $W$-axis. The zero temperature limit is considered in the picture.}
	\end{center}
\end{figure}

\begin{figure}
	\begin{center}
		\includegraphics[scale=0.58]{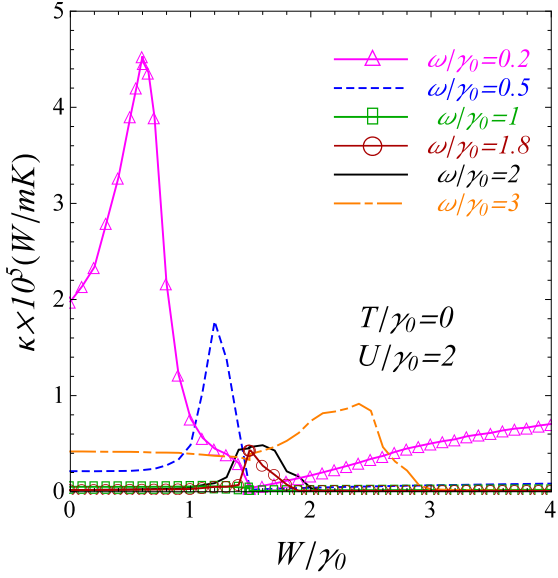}
		\caption{\label{fig:Fig_13}(Color online) The thermal conductivity $\kappa$ as a function of the interlayer Coulomb interaction parameter, for different values of the incident photon's frequencies. The zero temperature limit is considered in the picture.}
	\end{center}
\end{figure}

\begin{figure}
	\begin{center}
		\includegraphics[scale=0.58]{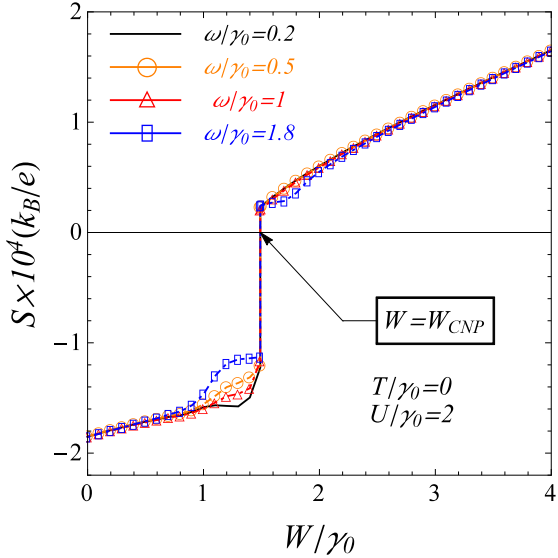}
		\caption{\label{fig:Fig_14}(Color online) The Seebeck coefficient $S$ as a function of the interlayer Coulomb interaction parameter, for different values of the incident photon's frequencies. The zero temperature limit is considered in the picture.}
	\end{center}
\end{figure}

\begin{figure}
	\begin{center}
		\includegraphics[scale=0.58]{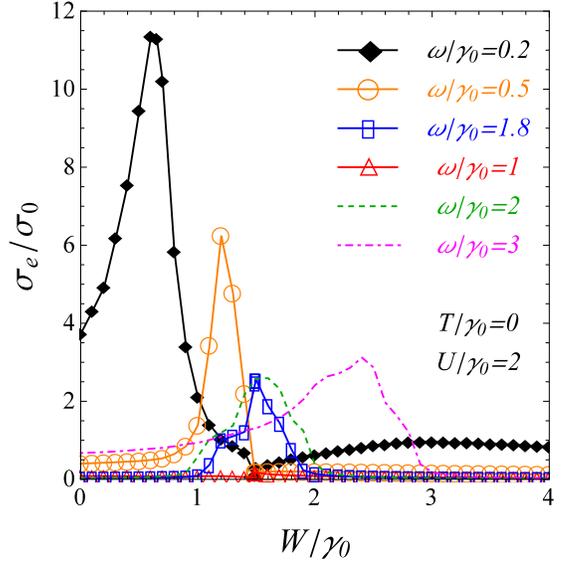}
		\caption{\label{fig:Fig_15}(Color online) The electrical conductivity as a function of the interlayer Coulomb interaction parameter, for different values of the incident photon's frequencies. The zero temperature limit is considered in the picture and the conductivity function is normalized at the elementary conduction quanta $\sigma_{0}=e^{2}/\hbar$.}
	\end{center}
\end{figure}

\begin{figure}
	\begin{center}
		\includegraphics[scale=0.58]{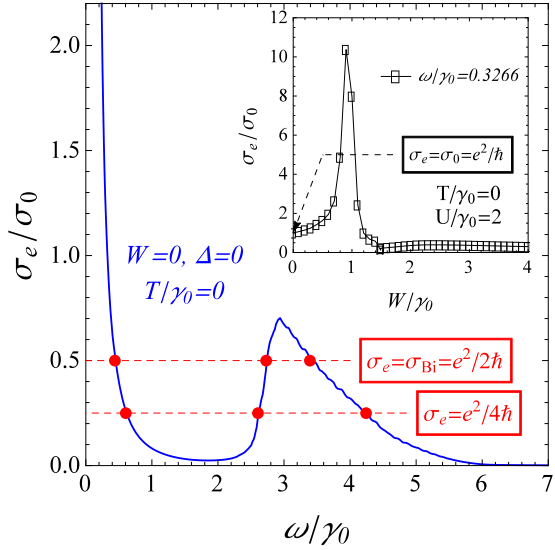}
		\caption{\label{fig:Fig_16}(Color online) The electrical conductivity, normalized to the elementary conductivity quanta $\sigma_{0}=e^{2}/\hbar$, as a function of the normalized photon's frequencies $\omega/\gamma_0$. The zero interaction limit is considered in the picture. Indicated red points in the picture correspond to different photon's energies at which the conductivity function gets the values equal to the minimal conductivities in the non-interacting monolayer ($\sigma_{\rm e}=e^{2}/4\hbar$) and bilayer ($\sigma_{\rm e}=e^{2}/2\hbar$) graphene systems. The temperature is set at zero.
		}
	\end{center}
\end{figure}

%
The thermoelectric FOM, given in Eq.(\ref{Equation_24}) can be rewritten as a function of the response functions as
\begin{eqnarray}    
ZT_{\rm e}=\frac{{\cal{L}}^{2}_{\rm 12}(\omega)}{{\cal{L}}_{\rm 11}(\omega){\cal{L}}_{\rm 22}(\omega)-{\cal{L}}^{2}_{\rm 12}(\omega)}.
\label{Equation_48} 
\end{eqnarray} 
The $\omega$ dependence of FOM is important to reveal out the optimal frequencies of the incident photons at which the FOM attains its maximal values. Theoretically, the thermoelectric figure of merit in bilayer graphene system has been discussed recently in Refs.\cite{cite_26, cite_29}. Particularly, the $\omega$-dependence of FOM has been considered in Ref.\cite{cite_26} concerning the AA-stacked bilayer graphene and its temperature dependence has been found for the biased structure. Here, we consider the effects of interactions and the excitonic pairing on the behavior of FOM. 
%
\begin{figure}
	\begin{center}
		\includegraphics[scale=0.58]{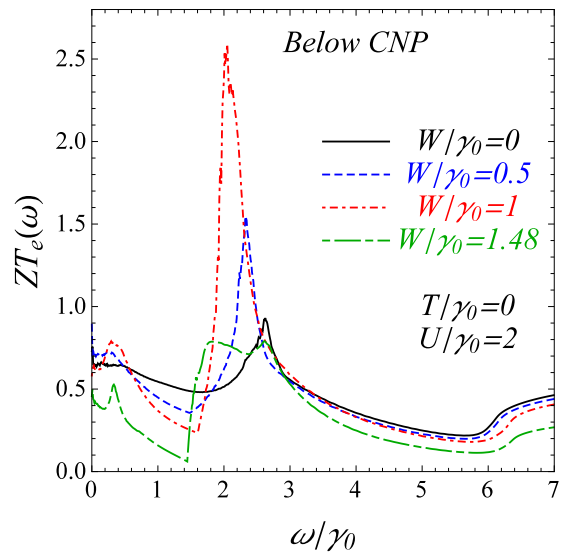}
		\caption{\label{fig:Fig_6}(Color online) The frequency dependence of electronic thermoelectric figure of merit (FOM), given in Eq.(\ref{Equation_48}). The considered values of the local interaction parameter $W$ are in the interval below the charge neutrality point (CNP) in the bilayer graphene.}
	\end{center}
\end{figure} 
%
Our treatment goes far beyond the standard tight-binding approaches discussed in Refs.\cite{cite_26, cite_29} and permits to consider properly the realistic thermoelectric application mechanisms for the bilayer graphene. In Fig.~\ref{fig:Fig_6}, we have presented the results for FOM for the same values of the interaction parameter $W$ as given in Fig.~\ref{fig:Fig_3}. It is clear that at the certain values of photon's frequencies the values of FOM are very high. The highest FOM value $ZT_{\rm e}=2.592$ is attained for $W=\gamma_0$ and at $\omega=2.05\gamma_0$. The $\omega$-dependence of FOM, for $W\geq W_{\rm CNP}$, is given in Fig.~\ref{fig:Fig_7}. It is clear from the results in Figs.~\ref{fig:Fig_6} and \ref{fig:Fig_7} that the highest FOM values are attained in the region $W<W_{\rm CNP}$. The positions of maxima of curves in Figs.~\ref{fig:Fig_6} and \ref{fig:Fig_7} don't change when increasing the temperature up to room temperature region $0\leq T \leq 300$ K (not presented here). The frequency dependence of the Seebeck coefficient is presented in Fig.~\ref{fig:Fig_8}. We have shown the curves which correspond to (from bottom to top) $W=0, 0.5\gamma_0, \gamma_0, 1.48\gamma_0, 1.4899\gamma_0, 2\gamma_0, 3\gamma_0, 4\gamma_0$. There are two curves in Fig.~\ref{fig:Fig_8} corresponding to the critical value $W_{\rm CNP}=1.4899\gamma_0$. One of them (with the lower bound solution for $\mu$: $\mu=\mu_{\rm LB}$) is in the negative region $S(\omega)<0$, while the other one (with the upper bound solution for $\mu$: $\mu=\mu_{\rm UB}$) is in the positive region $S(\omega)>0$. They define, indeed, the outermost boarders of the coefficient $S(\omega)$ at which the it changes its sign. We see also in Fig.~\ref{fig:Fig_8} that the Seebeck coefficient does not vary considerably with the photon's frequencies $\omega$.  
%
%
We see clearly in that the Seebeck coefficient changes its sign when crossing the CNP point (see the curves in dark yellow and orange, in Fig.~\ref{fig:Fig_8}, corresponding the lower and upper bounds of the chemical potential $\mu$). It has been shown experimentally by Yuri M. Zuev, et al., \cite{cite_52}, that the sign of the TEP changes across the charge neutrality point as the majority carrier density switches from electron to hole. This result is consistent with our behavior of the Seebeck coefficient, given in Fig.~\ref{fig:Fig_8}. The similar effects in thermopower have been shown also in Refs.\cite{cite_17, cite_18}. The temperature dependence of the FOM and the electrical conductivity function $\sigma_{\rm e}(T)$ are shown in Figs.~\ref{fig:Fig_9}, \ref{fig:Fig_10} and \ref{fig:Fig_11}. Different interaction limits are considered in the pictures. In Fig.~\ref{fig:Fig_10}, the electrical conductivity function $\sigma_{\rm e}$ is normalized to the minimum conductivity quanta $\sigma_{0}=e^{2}/\hbar$. We see in Fig.~\ref{fig:Fig_9} that at $W<W_{\rm CNP}$ the FOM values are very close to the value $ZT_{\rm e}=1$, for a very large interval of temperature, while for $W>W_{\rm CNP}$ FOM is decreasing rapidly when increasing the temperature (see the small-dashed curve in green and the solid curve in orange). The highest values of the electrical conductivity function are obtained at $W=\gamma_0$, which corresponds to the behavior of FOM presented in Figs.~\ref{fig:Fig_6} and \ref{fig:Fig_9}. It is worth to mention here that the external field frequency $\omega$, for each curve in Fig.~ \ref{fig:Fig_9}, is fixed to the value at which the FOM attains its maximal values in Figs.~\ref{fig:Fig_6} and \ref{fig:Fig_7} (for the case $T=0)$.  
%
%
The behavior of the thermoelectric figure of merit as a function of temperature, starting from zero up to room temperature value $T=300$ eV (corresponding to the normalized temperature $T/\gamma_0=0.0086$), is given in Fig.~\ref{fig:Fig_11} below. We see that the FOM is very stable to the variation of temperature and the considerable deviation from this behavior takes place at the value $W=2\gamma_0=6$ eV (see the small-dashed curve in green). The perfect thermoelectric FOM is achieved at $W=\gamma_0=3$ eV, at which very high values of FOM (given in the interval $2.4\leq ZT_{\rm e}\leq 2.592$) are obtained in the interacting BLG with excitons in the temperature interval $T\in[0,300]K$. This result suggests that the interacting bilayer graphene could be a high-FOM performance thermoelectric device, operating at the room temperatures. The electrical conductivity function $\sigma_{\rm e}$ is continuously increasing with $T$ in this range of temperature (see in Fig.~\ref{fig:Fig_10}) and contributes in the high values of $ZT_{\rm e}$ (see in Eq.(\ref{Equation_24})) and we will see that the principal shape of thermoelectric conductivity $\kappa$ is due to the electrical conductivity function $\sigma_{\rm e}$. The difficult experimental task, related to the observation of our results is related to the finding of appropriate values of external field frequencies at which the BLG will be calibrated to the suitable interaction value $W$. This could be achieved with the help of the adiabatic variation of the external field values (see in Ref.\cite{cite_53}, where such a situation is described in the context of the excitonic Josephson tunnel junction based on AB-BLG) until the suitable effect takes place in the BLG.   
%
%
%
%
It is also interesting to discuss the explicit $W$-dependence of thermal parameters in the interacting AB-BLG in order to observe the changes of those parameters at the CNP point and finding the similarities in their behavior. In Figs.~\ref{fig:Fig_12}, ~\ref{fig:Fig_13}, ~\ref{fig:Fig_14}, and ~\ref{fig:Fig_15}, we shown the behavior of principal thermoelectric quantities defined in Eq.(\ref{Equation_28}). In Fig.~\ref{fig:Fig_12}, the FOM is given for $W$ in the interval $W\in [0,4\gamma_0]$. This is a large enough interval that covers the small and strong interaction limits in the AB-BLG. Then we have considered several values of frequencies $\omega$, from low ($\omega=0.1\gamma_0=0.3$ eV) up to very high frequencies ($\omega \sim 3\gamma_0=9$ eV). We see in Fig.~\ref{fig:Fig_12} that for all fixed values of the photon's frequencies $\omega$ the thermoelectric FOM parameter has a drastic decrease at the value $W=W_{\rm CNP}$, corresponding to the charge neutrality point. It is clear in Fig.~\ref{fig:Fig_12} that the largest FOM values are given in the interval of Coulomb interaction $W\in(0, W_{\rm CNP})$, i.e., for $W$ below the CNP value and for which the AB-BLG system is in the mixed state composed of the excitonic insulator and condensate states (see in Ref.\cite{cite_38, cite_41}). The highest FOM values are attained at $\omega=1.8\gamma_0=5.4$ eV and in the interval $W\in[1.12\gamma_0,1.4\gamma_0]$ (or $W\in[3.36, 4.2]$ eV), given by the solid line in red in Fig.~\ref{fig:Fig_12}. In this interval of the interaction parameter the FOM $ZT_{\rm e}$ takes the values $ZT_{\rm e}\in[2.164, 2.865]$ , with the highest value $ZT^{{\rm max}}_{\rm e}=2.865$ at $W_{0}=1.33\gamma_0=3.99 eV$. The critical value of $W$ at $W=1.4988\gamma_0=4.47 eV$ is indicated by the arrow, in Fig.~\ref{fig:Fig_12}. The behavior of thermal conductivity $\kappa$ as a function of interaction parameter $W$ is given in Fig.~\ref{fig:Fig_13}.  	
%
%
Contrary, the parameter $\kappa$ takes the small values in the interval where $ZT_{\rm e}$ is high (this is consistent with the equation in Eq.(\ref{Equation_24}), where the large values of $\kappa$ suggest the small values of $ZT_{\rm e}$). We see in Fig.~\ref{fig:Fig_13} that when increasing $\omega$ in the interval $\omega\in(0, \gamma_0)$ the function $\kappa$ decreases in the region of the interaction parameter $W\in[0,W_{\rm CNP}]$ and takes very small values at $\omega=\gamma_0$ (see the curve in green, in Fig.~\ref{fig:Fig_13}). The thermal conductivity vanishes at the CNP point for the photon's frequencies $\omega\in [0.2\gamma_0,\gamma_0]$, while it is finite at the CNP for the frequencies in the interval $\omega\in[1.8\gamma_0,3\gamma_0]$.
%
%
%
In Fig.~\ref{fig:Fig_14}, we have shown the critical behavior of the Seebeck coefficient $S$ as a function of $W$. It is clear in Fig.~\ref{fig:Fig_14} that the coefficient $S$ reflects well the particle-hole symmetry in our BLG system and changes the sign exactly at the CNP point given by $W=W_{\rm CNP}$. Moreover, the Seebeck coefficient does not vary when changing the frequencies and only small changes happen in the electron-region $S<0$, near the CNP point. This is due to the fluctuations of the chemical potential in the deep vicinity of the CNP point. 
%
%
The behavior of the electrical conductivity $\sigma_{\rm e}$ as a function of $W$ is given in Fig.~\ref{fig:Fig_15} and it is very similar to that of the electronic thermal conductivity function $\kappa$, given in Fig.~\ref{fig:Fig_13}. As it was mentioned earlier, it repeats the shapes of the curves of $\kappa$ in Fig.~\ref{fig:Fig_13} but with different scales. This suggests that in the limit $T\rightarrow 0$ the Wiedemann-Franz law should be satisfied. It is worth to mention that the electrical conductivity function presented here (as a consequence of temperature gradient in the system) differs much from the optical conductivity function discussed in Ref.\cite{cite_40}, where a threshold behavior is observed in the conductivity spectrum. We see in Fig.~\ref{fig:Fig_15} that the electrical conductivity in the AB-BLG system is not a threshold process and the function $\sigma_{\rm e}$ is non zero for all values of $W$, except the charge neutrality point, i.e., $W=W_{\rm CNP}$. Another interesting behavior of the electrical conductivity function $\sigma_{\rm e}(\omega)$ is related to the case of the non-interacting AB-BLG system. This case is presented here, in Fig.~\ref{fig:Fig_16}. The conductivity limit corresponding to the non-interacting monolayer graphene $\sigma_{\rm e}=\sigma_{\rm MG}=e^{2}/4\hbar$ could be achieved for three different values of the external photon's frequencies: low-frequency limit $\omega_{01}=0.617\gamma_0$, medium frequency limit $\omega_{02}=2.618\gamma_0$ and high-frequency limit with $\omega_{03}=4.254\gamma_0$. The electrical conductivity corresponding to the non-interacting bilayer graphene with the minimal electrical conductivity $\sigma_{\rm e}=\sigma_{\rm Bi}=e^{2}/2\hbar$ \cite{cite_54, cite_55, cite_56} could be achieved also at three different values of the photon's energies. Namely, at $\omega_{01}=0.452\gamma_0$ (low-frequency), $\omega_{02}=2.742\gamma_0$ (medium-frequency) and $\omega_{03}=3.404\gamma_0$ (high-frequency). Those points are indicated on the curve of the normalized electrical conductivity function in Fig.~\ref{fig:Fig_16}. The inset, in Fig.~\ref{fig:Fig_16} shows the interaction dependence of the electrical conductivity function. The frequency $\omega/\gamma_0= 0.3266$ (or $\omega=0.97$ eV) is chosen such that $\sigma_{\rm e}=\sigma_{0}$ in the limit of the non-interacting bilayer $W/\gamma_0=0$.  
%
%
\begin{figure}
	\begin{center}
		\includegraphics[scale=0.58]{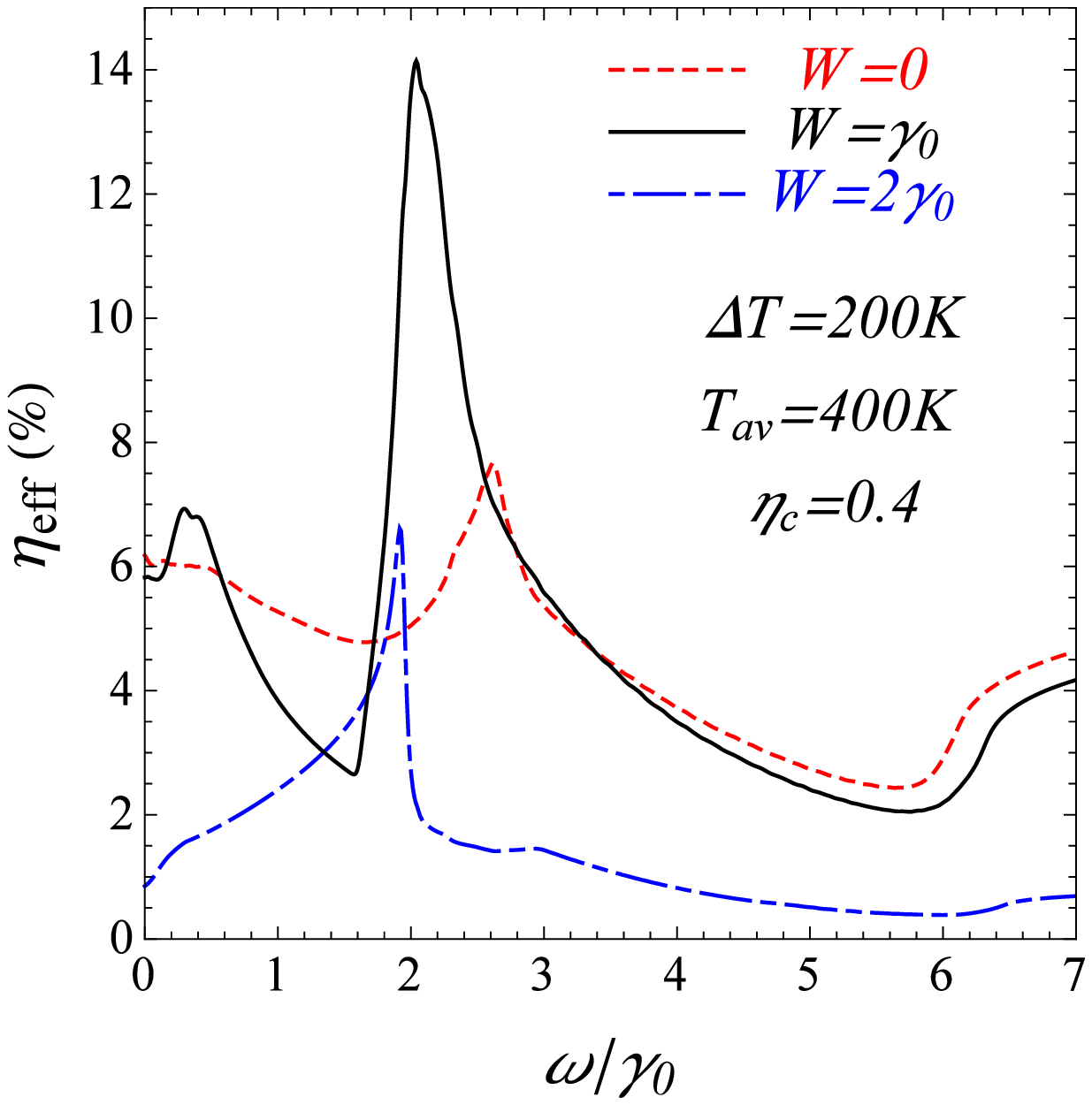}
		\caption{\label{fig:Fig_17}(Color online) Thermoelectric efficiency parameter $\eta_{\rm eff}$ in AB bilayer graphene, given in Eq.(\ref{Equation_25}), as a function of the normalized photon's frequencies $\omega/\gamma_0$. The temperatures $T_{\rm c}=300$ K and $T_{\rm h}=500$ K are assumed at the cold and hot sides of the system. Different values of the interlayer Coulomb interaction parameter $W$ are considered in the picture.}
	\end{center}
\end{figure} 
%
\begin{figure}
	\begin{center}
		\includegraphics[scale=0.58]{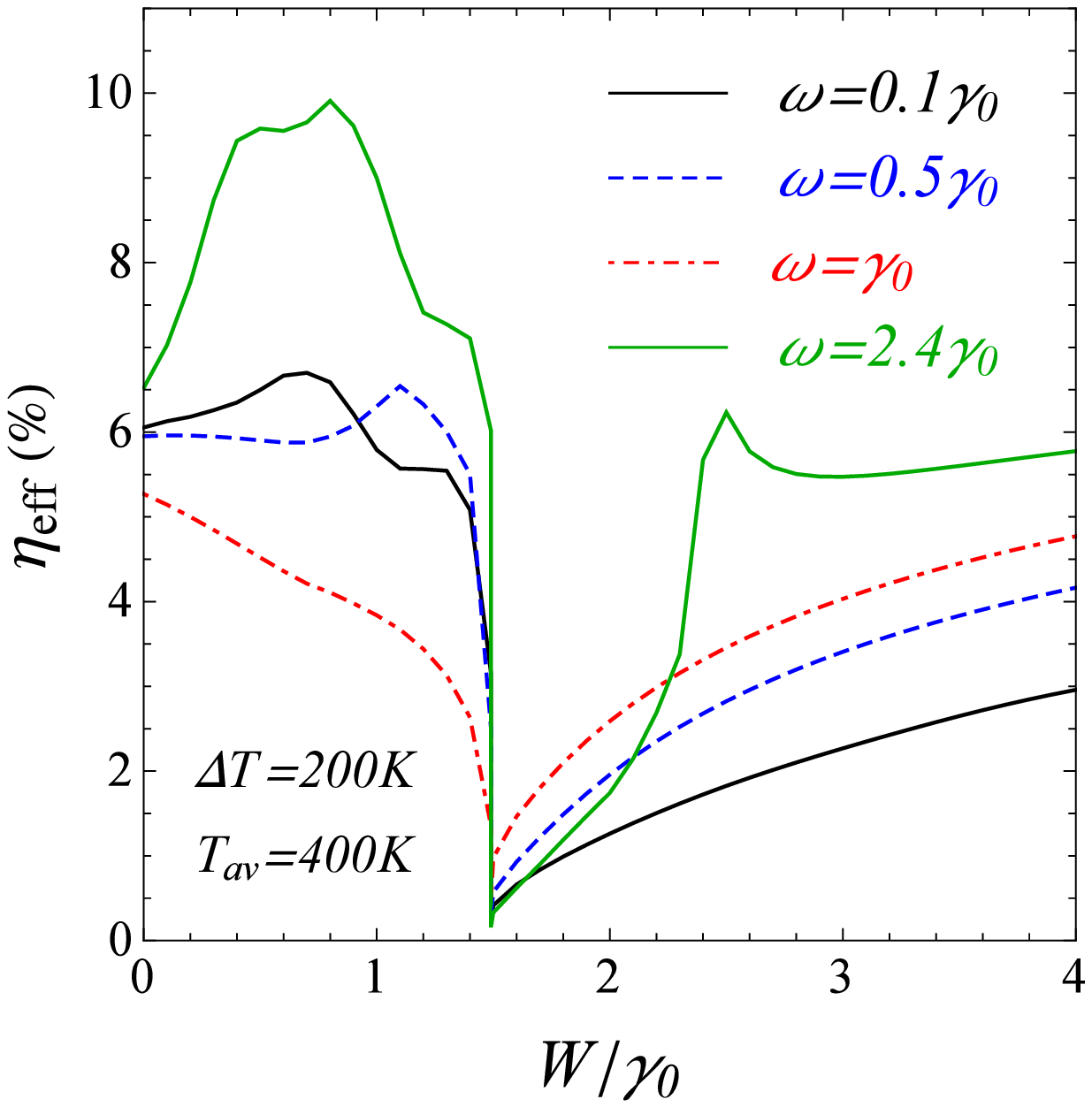}
		\caption{\label{fig:Fig_18}(Color online) Thermoelectric efficiency parameter $\eta_{\rm eff}$ in AB bilayer graphene, given in Eq.(\ref{Equation_25}), as a function of the normalized Coulomb interaction parameter $W/\gamma_0$ in the AB-BLG. The temperatures $T_{\rm c}=300$K and $T_{\rm h}=500$K are assumed at the cold and hot sides of the system. Different values of the normalized photon's frequencies are considered in the picture.}
	\end{center}
\end{figure} 
%
\begin{figure}
	\begin{center}
		\includegraphics[scale=0.58]{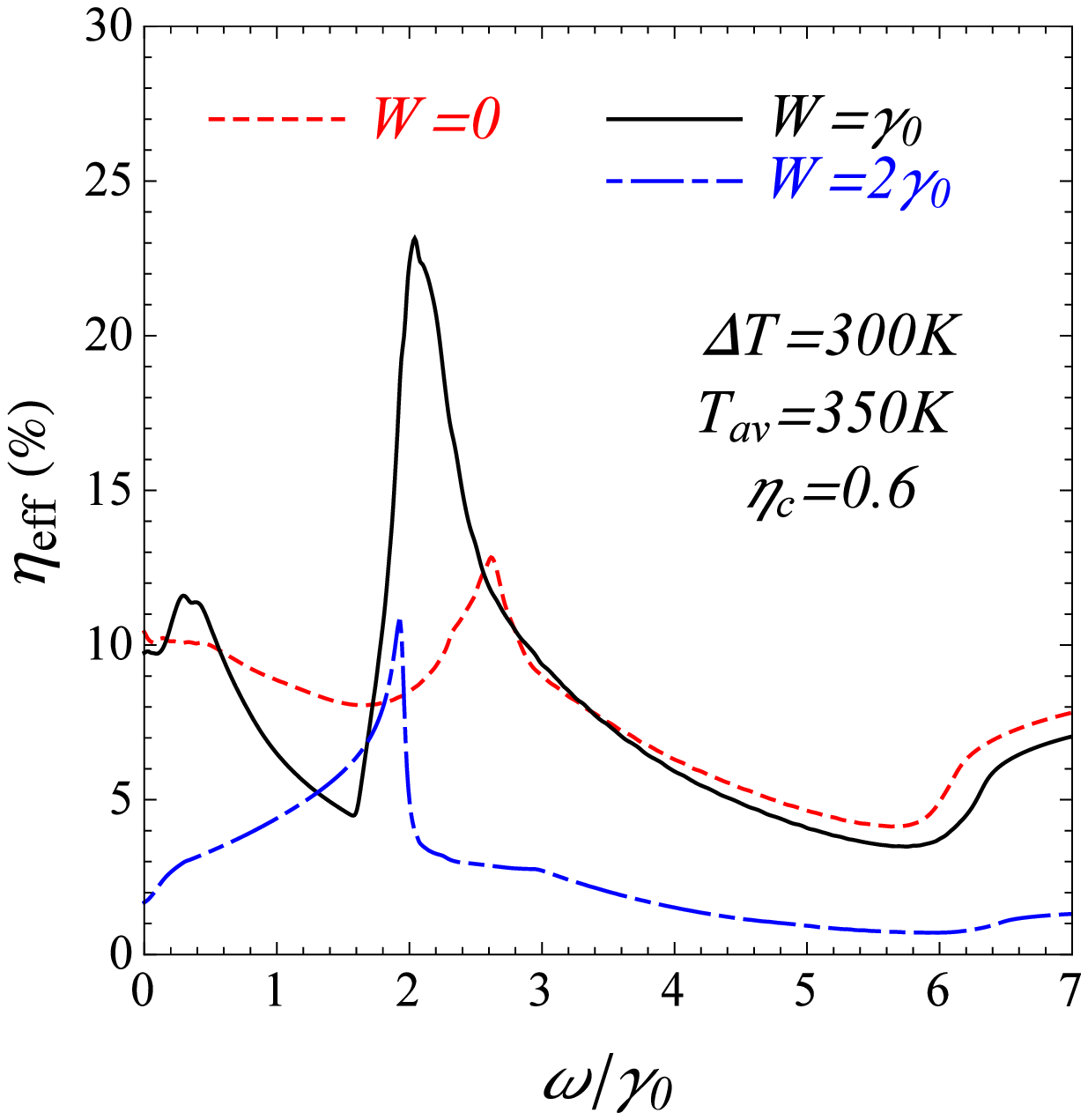}
		\caption{\label{fig:Fig_19}(Color online) Thermoelectric efficiency parameter $\eta_{\rm eff}$ in AB bilayer graphene, given in Eq.(\ref{Equation_25}), as a function of the normalized photon's frequencies $\omega/\gamma_0$. The temperatures $T_{\rm c}=200$ K and $T_{\rm h}=500$ K are assumed at the cold and hot sides of the system. Different values of the interlayer Coulomb interaction parameter $W$ are considered in the picture.}
	\end{center}
\end{figure} 
%
\begin{figure}
	\begin{center}
		\includegraphics[scale=0.58]{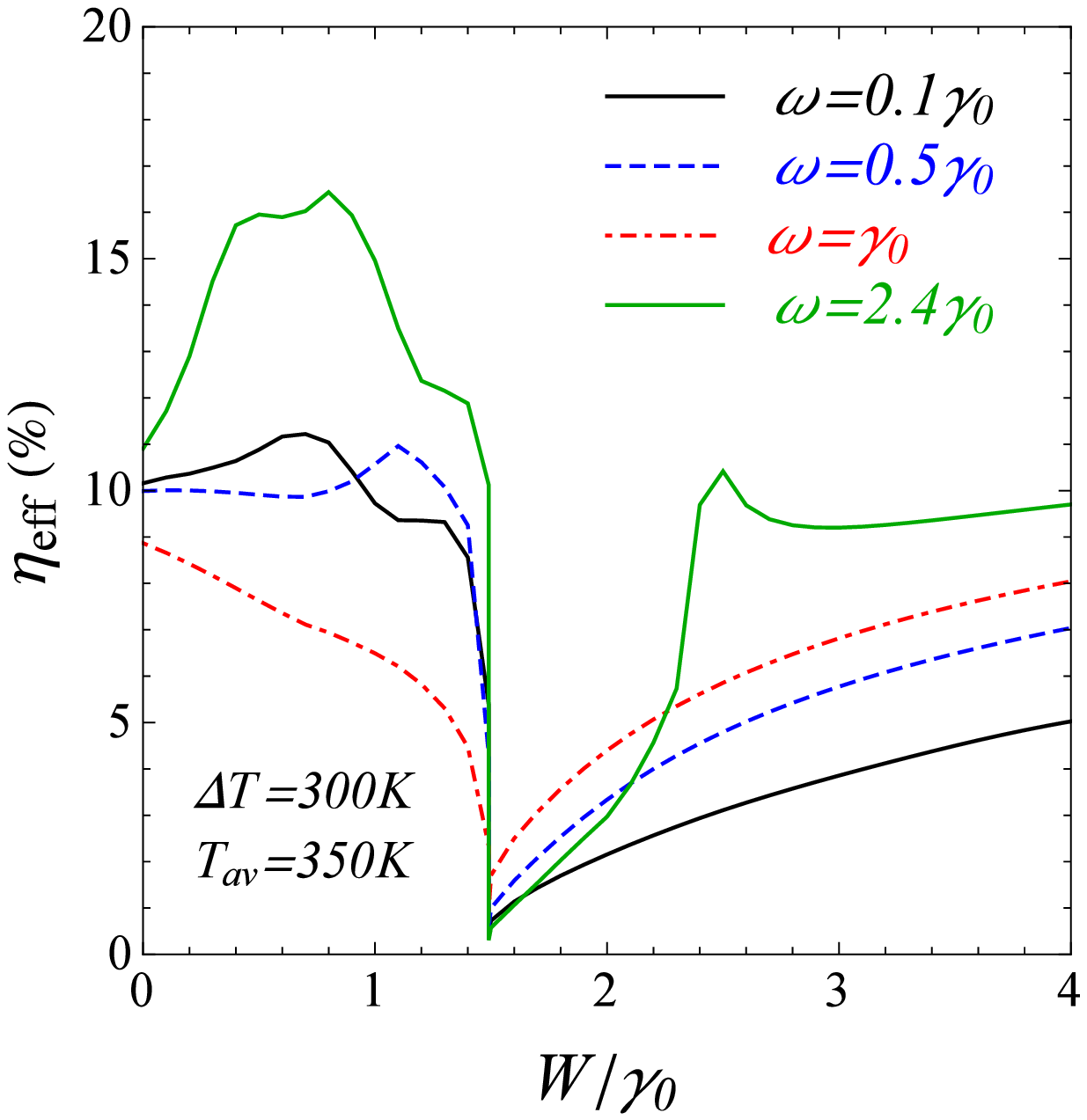}
		\caption{\label{fig:Fig_20}(Color online) Thermoelectric efficiency parameter $\eta_{\rm eff}$ in AB bilayer graphene, given in Eq.(\ref{Equation_25}), as a function of the normalized Coulomb interaction parameter $W/\gamma_0$ in the AB-BLG. The temperatures $T_{\rm c}=200$K and $T_{\rm h}=500$K are assumed at the cold and hot sides of the system. Different values of the normalized photon's frequencies are considered in the picture.}
	\end{center}
\end{figure} 
%
\subsection{\label{sec:Section_5_3} Thermoelectric efficiency}
%
In general, it is known that higher average FOM ${ZT}_{\rm av}$ and larger temperature difference $\Delta{T}$ will produce the higher energy conversion efficiency $\eta_{\rm eff}$. For the traditional heat engines which rival the Carnot efficiency $\eta_{\rm c}=\Delta{T}/T_{\rm h}$ with $\Delta{T}=400$ K and ${ZT}_{\rm av}\approx 3$ we have the thermoelectric conversion efficiency $\eta_{\rm eff}\sim 25 \%$ \cite{cite_57, cite_58}. Here, we calculate the thermoelectric conversion efficiency coefficient $\eta_{\rm eff}$ for two different regimes of the Carnot efficiency. Namely, $\eta_{\rm c}=0.4$ which corresponds to temperature difference $\Delta{T}=200$ K with $T_{\rm c}=300$ K and $T_{\rm h}=500$ K and $\eta_{\rm c}=0.6$ which corresponds to temperature difference $\Delta{T}=300$ K with $T_{\rm c}=200$ K and $T_{\rm h}=500$ K. In Fig.~\ref{fig:Fig_17}, we have shown the $\omega$-dependence of the coefficient $\eta_{\rm eff}$ (in $\%$) for the first case. Different values of $W$ are considered in the picture. We see that for $W\leq W_{\rm CNP}$ the coefficient $\eta_{\rm eff}$ is very high in a given interval of the photon's frequencies, situated in the deep ultraviolet and vacuum ultraviolet range of the spectrum. Particularly, when $\omega\in(1.84\gamma_0, 2.6\gamma_0)$ the values of the coefficient $\eta_{\rm eff}$ are in the range $\eta_{\rm eff}\in(7, 14.14)\%$. Contrary, at $W=2\gamma_0$ (see the large dot-dashed curve blue, in Fig.~\ref{fig:Fig_17}) the coefficient $\eta_{\rm eff}$ is reduced considerably. Thus, at $W=\gamma_0$ and for $\omega\in(5.52,7.8)$ eV the thermal performance of the interacting AB-BLG is very high (the solid black curve in Fig.~\ref{fig:Fig_17}). In Fig.~\ref{fig:Fig_18}), we have presented the $W$-dependence of the coefficient $\eta_{\rm eff}$ with $\eta_{\rm c}=0.4$ and for different values of the photon's frequencies $\omega$. The general behavior of $\eta_{\rm eff}$ in Fig.~\ref{fig:Fig_18} is governed by the thermoelectric FOM, presented in Fig.~\ref{fig:Fig_12}, in the Section \ref{sec:Section_5_2}. Namely, there is a drastic decrease of $\eta_{\rm eff}$ at the charge neutrality point $W=W_{\rm CNP}$ and the high values of $\eta_{\rm eff}$ correspond to the high energy photons (see the solid green line, in Fig.~\ref{fig:Fig_18}) with $\omega=2.4\gamma_0=7.2$ eV). The thermoelectric coefficient enhances considerably when increasing the Carnot efficiency to the value $\eta_{\rm c}=0.6$ (see in Figs.~\ref{fig:Fig_19} and ~\ref{fig:Fig_20} here). We observe in Fig.~\ref{fig:Fig_19} that for $W<W_{\rm CNP}$ (see the curves at $W=0$ and $W=\gamma_0$) the coefficient $\eta_{\rm eff}$ is higher than $5\%$ practically for all values of the external photon's energies. In the narrow regions on $\omega$-axis ($\omega\in(0,0.495\gamma_0)=(0,1.485)$ eV) and $\omega\in(2.3\gamma_0, 2.83\gamma_0)=(6.9,8.49)$ eV, we have $\eta_{\rm eff}{\gtrsim} 10\%$ for the case $W=0$ (see the dashed red curve in Fig.~\ref{fig:Fig_19}). For $W=\gamma_0$ (see the solid black curve in Fig.~\ref{fig:Fig_19}) we have $\eta_{\rm eff}\gtrsim 10\%$ for the photon's energies in the intervals $\omega\in(0.161\gamma_0,0.545\gamma_0)=(0.483,1.635) eV$ (low-energy photons) and $\omega\in(1.787\gamma_0, 2.87\gamma_0)=(5.361,8.61)$ eV (high-energy photons). There is also a very narrow and high energy interval $\omega\in(1.95\gamma_0,2.23\gamma_0)=(5.85,6.69)$ eV, where we have $\eta_{\rm eff}\gtrsim 20\%$. For the $W$-dependence of $\eta_{\rm eff}$, in this case, we have the same behavior as for the previous case with $\eta_{\rm c}=0.4$, but the values of $\eta_{\rm eff}$ are much higher. The obtained here values for the energy conversion efficiency coefficient permit to qualify definitively the bilayer graphene among the high performance thermoelectric materials and open a new prospect towards the usage of bilayer graphene as a material with outstanding thermoelectric properties.   
%
\section{\label{sec:Section_6} Conclusion}
%
We considered here the thermoelectric properties in the excitonic BLG system with the Coulomb interaction effects, included in the band structure dispersion relation. By supposing the temperature gradient in the individual graphene-layers in the AB-stacked BLG, we considered the electrical and heat current components along the $x$-axis in the two-dimensional layers. Then after using the Mahan's method \cite{cite_44} for the electronic and heat currents density operators, we evaluated the explicit analytical forms of the linear response coefficients within the Kubo-Green formalism. The principal thermoelectric parameters have been calculated in the excitonic BLG and the role of the charge neutrality point has been discussed. We obtained the additional single peaks (due to the interlayer excitons) in the spectrum of the thermal conductivity function $\kappa$, for the values of the interaction parameter $W$ in the region $W<W_{\rm CNP}$ and in the low-frequency regime $\omega\in[0,\gamma_0]$. While for $W\geq W_{\rm CNP}$ we get the usual behavior for the conductivity function in the low-frequency regime (see for example in Ref.\cite{cite_23}). The observed double-peak structure in the low-frequency $\kappa$ spectrum and the additional peaks in the high photon's frequency region are attributed to the existence of the excitonic insulator state in the BLG \cite{cite_38} and the existence of the excitonic condensate state in the AB-BLG, even for the case $W=0$ \cite{cite_41}. We have calculated the Seebeck coefficient and electrical conductivity function $\sigma$ normalized to the minimal conductivity quanta $e^{2}/\hbar$. We have shown that the thermoelectric figure of merit (FOM) could hit very high values ($ZT\gtrsim 2.5$) in the excitonic BLG, for a given interaction parameter regime and a given interval of the photon's energy. Additionally, we have calculated the temperature dependence of the FOM for different Coulomb interaction regimes and we found nearly a constant picture up to room temperatures suggesting the high stability of FOM with respect to temperature changes and we have shown the possibility of room temperature applications of our BLG as an efficient thermoelectric engine. Another remarkable effect is related to the electrical conductivity mediated by the temperature gradient in the BLG. Indeed, we have shown that the limits of the minimal electrical conductivities $e^{2}/4\hbar$ and $e^{2}/2\hbar$ typical for the non-interacting monolayer and bilayer graphene systems could be reached at three different operating frequencies for our BLG system (at $W=0$ or $\Delta=0$). 
Moreover, we found that thermopower changes the sign at the charge neutrality point $W=W_{\rm CNP}$ by defining the branches for electrons ($S<0$, at $W<W_{\rm CNP}$) and holes ($S>0$, at $W>W_{\rm CNP}$), thus, behaving in accordance with the definition of the excitonic insulator state, given in Ref.\cite{cite_38}. The very high values of the thermoelectric conversion efficiency coefficient obtained here ($\eta_{\rm eff}\gtrsim 25\%$, for not very large temperature differences $\Delta{T}$), qualifies the excitonic BLG as a high-performance thermoelectric energy conversion material and open a new prospect for the practical technological applications of the excitonic AB-BLG such as in the modern solar cell energy converters, thermoelectric nano-converters for the effective heating or cooling the microprocessors, the heat converter engines for the ecological locomotions, and etc.. 

At the end, we have to distinguish, among many cited extraordinary thermal properties in the AB-BLG, two principal advantages for which the thermoelectric applications of such a system are strongly recommended. 
First one, is related to the high values of $ZT$ ($ZT\sim 2.5$) at the room temperatures which suggest the possibility of direct device applications of the excitonic BLG system as a high-performance thermoelectric material operating at the room temperatures. The high values of the thermal conversion efficiency ($\eta_{\rm eff}\gtrsim 25\%$), from another hand, imply that the AB bilayer graphene system is a new efficient thermoelectric energy conversion device with the unique excitonic band structure.
In our suggestion, the excitonic bilayer graphene could be reliable also as a model system for the upcoming thermometry devices at the nanoscale and the new generation laser sensors.

The authors have no competing interests to declare

\appendix
%
\section{\label{sec:Section_7} Mahan's formalism}
%
\subsection{\label{sec:Section_7_1} Electric current density operator ${\bf{j}}_{\rm e}$}
%
The equation of continuity for the electric current density operator in the Matsubara time formalism reads as 
\begin{eqnarray}    
\frac{\partial n(\tau)}{\partial \tau}-i\nabla{{\bf{j}}_{\rm e}}=0,
\label{Equation_7_1} 
\end{eqnarray}
where $n(\tau)$ is the total electron density operator in the bilayer graphene $n(\tau)=n_{1}(\tau)+n_{2}(\tau)$. The electric polarization operator, introduced in the Section \ref{sec:Section_3} is related to the electron density operator $n$ by ${\bf{P}}_{\rm e}=\sum_{{\bf{m}}}{\bf{R}}_{\bf{m}}n_{{\bf{m}}}$, where the summation is over the lattice sites positions ${\bf{m}}$, and ${\bf{R}}_{{\bf{m}}}$ is the lattice translation vector at the site ${\bf{m}}$. We have ${\bf{j}}_{\rm e}=i\partial{\bf{P}}_{\rm e}/\partial{\tau}=i\left[H,{\bf{P}}_{\rm e}\right]$, where we have used the Heisenberg equation of motion for the operator ${\bf{P}}_{\rm e}$. Then, the Hamiltonian of the bilayer graphene system, given in Eq.(\ref{Equation_1}) can be written as a sum of the site-dependent effective Hamiltonian density operator $h_{\rm eff{\bf{m}}}$, i.e., $H=\sum_{{\bf{m}}}h_{\rm eff{\bf{m}}}$, after which we get the result given in Eq.(\ref{Equation_29}), where only the $x$-component of the operator ${\bf{j}}_{\rm e}$ is considered 
\begin{eqnarray}    
{\bf{j}}_{\rm e}=i\sum_{{\bf{m}},{\bf{n}}}{\bf{R}}_{{\bf{m}}}\left[h_{\rm eff{\bf{n}}},n_{\bf{m}}\right].
\label{Equation_7_2} 
\end{eqnarray}
Thus, the calculation of the electric current density operator ${\bf{j}}_{\rm e}$ reduces to the calculation of the standard commutator in the right-hand side in Eq.(\ref{Equation_7_2}). The effective Hamiltonian density operator $h_{\rm eff{\bf{n}}}$ after the linearization procedure described in Ref.\cite{cite_38} is
\begin{eqnarray}    
h_{\rm eff{\bf{m}}}&&=-\gamma_0\sum_{{\bm{\mathit{\delta}}}_{1}\sigma}\left(a^{\dag}_{1,{\bf{m}}+{\bm{\mathit{\delta}}}_{1}\sigma}b_{1,{\bf{m}}\sigma}+b^{\dag}_{1,{\bf{m}+{\bm{\mathit{\delta}}}_{1}}\sigma}a_{1{\bf{m}}\sigma}\right)
\nonumber\\
&&-\gamma_0\sum_{{\bm{\mathit{\delta}}}_{2}\sigma}\left(a^{\dag}_{2,{\bf{m}}+{\bm{\mathit{\delta}}}_{2}\sigma}b_{2,{\bf{m}}\sigma}+b^{\dag}_{2,{\bf{m}+{\bm{\mathit{\delta}}}}_{2}\sigma}a_{2,{\bf{m}}\sigma}\right)
\nonumber\\
&&-\left(\gamma_{1}+\Delta\right)\sum_{\sigma}\left(a^{\dag}_{2,{\bf{m}}\sigma}b_{1,{\bf{m}}\sigma}+b^{\dag}_{1,{\bf{m}}\sigma}a_{2{\bf{m}}\sigma}\right)
\nonumber\\
&&-\mu_{1\rm eff}\sum_{\sigma}\left(n_{1a_{1}{\bf{m}}}+n_{2b_{2}{\bf{m}}}\right)
\nonumber\\
&&-\mu_{2\rm eff}\sum_{\sigma}\left(n_{2a_{2}{\bf{m}}}+n_{1b_{1}{\bf{m}}}\right).
\label{Equation_7_3} 
\end{eqnarray}
Here, the parameter $\Delta$ appears after the linearization of the interlayer interaction term in Eq.(\ref{Equation_1}), while the effective chemical potentials $\mu_{\rm 1eff}$ and $\mu_{\rm 2eff}$, defined in the Section \ref{sec:Section_2_1}, result from the linearization of the intralayer Hubbard interaction term $U$ and the interlayer interaction term $W$. 
The total electron density operator $n_{\bf{m}}$, in the commutator in Eq.(\ref{Equation_7_2}), is $n_{\bf{m}}=\sum_{\ell,\sigma}n_{\ell,{\bf{m}}\sigma}$, where the summation index $\ell=1,2$ denotes the layers in the bilayer graphene. Then, for the layer $\ell=1$ we get from Eq.(\ref{Equation_7_2}) 
\begin{eqnarray}    
&&{\bf{j}}^{\ell=1}_{\rm e}=-\gamma_0\sum_{{\bf{n}}{\bf{m}}}\sum_{{\bm{\mathit{\delta}}}_{1}\sigma}{\bf{R}}_{m}\left(-\delta_{{\bf{n}}{\bf{m}}}a^{\dag}_{1,{\bf{m}}\sigma}b_{1,{\bf{n}}+{\bm{\mathit{\delta}}}_{1}\sigma}\right.
\nonumber\\
&&\left.+\delta_{{\bf{m}}{\bf{n}}+{\bm{\mathit{\delta}}}_{1}}a^{\dag}_{1,{\bf{n}}\sigma}b_{1,{\bf{m}}\sigma}+\delta_{{\bf{m}}{\bf{n}}}b^{\dag}_{1,{\bf{n}}+{\bm{\mathit{\delta}}}_{1}\sigma}a_{1,{\bf{m}}\sigma}\right.
\nonumber\\
&&\left.-\delta_{{\bf{m}}{\bf{n}}+{\bm{\mathit{\delta}}}_{1}}b^{\dag}_{1,{\bf{m}}\sigma}a_{1,{\bf{n}}\sigma}
\right)
\label{Equation_7_4} 
\end{eqnarray}
and for the layer $\ell=2$ we have
\begin{eqnarray}    
&&{\bf{j}}^{\ell=2}_{\rm e}=-\gamma_0\sum_{{\bf{n}}{\bf{m}}}\sum_{{\bm{\mathit{\delta}}}_{2}\sigma}{\bf{R}}_{m}\left(-\delta_{{\bf{n}}{\bf{m}}}a^{\dag}_{2,{\bf{m}}\sigma}b_{2,{\bf{n}}+{\bm{\mathit{\delta}}}_{2}\sigma}\right.
\nonumber\\
&&\left.+\delta_{{\bf{m}}{\bf{n}}+{\bm{\mathit{\delta}}}_{2}}a^{\dag}_{2,{\bf{n}}\sigma}b_{2,{\bf{m}}\sigma}+\delta_{{\bf{m}}{\bf{n}}}b^{\dag}_{2,{\bf{n}}+{\bm{\mathit{\delta}}}_{2}\sigma}a_{1,{\bf{m}}\sigma}\right.
\nonumber\\
&&\left.-\delta_{{\bf{m}}{\bf{n}}+{\bm{\mathit{\delta}}}_{1}}b^{\dag}_{2,{\bf{m}}\sigma}a_{1,{\bf{n}}\sigma}
\right).
\label{Equation_7_5} 
\end{eqnarray}
Combining the results in Eqs.(\ref{Equation_7_4}) and (\ref{Equation_7_5}) we obtain for the total electron current density operator ${\bf{j}}_{\rm e}={\bf{j}}^{\ell=1}_{\rm e}+{\bf{j}}^{\ell=2}_{\rm e}$ the following result
\begin{eqnarray}    
{\bf{j}}_{\rm e}=-\gamma_0\sum_{{\bf{n}}\ell\sigma{\bm{\mathit{\delta}}}_{\ell}}{\bm{\mathit{\delta}}}_{\ell}\left( a^{\dag}_{\ell,{\bf{n}}\sigma}b_{\ell,{\bf{n}}+{\bm{\mathit{\delta}}}_{\ell}\sigma}-b^{\dag}_{\ell,{\bf{n}}+{\bm{\mathit{\delta}}}_{\ell}\sigma}a_{\ell,{\bf{n}}\sigma}\right).
\nonumber\\
\label{Equation_7_6} 
\end{eqnarray}
It is not difficult to show that the commutator of the total electron density operator with the interlayer part of the Hamiltonian density, given in Eq.(\ref{Equation_7_3}) is zero. Indeed, we have 
\begin{eqnarray}    
\left[h_{{\bf{n}}}(\gamma_1),n_{{\bf{m}}}\right]=\left[h_{{\bf{n}}}(\gamma_1),\sum_{\sigma'\ell}{\bf{R}}_{{\bf{m}}}\left(a^{\dag}_{\ell,{\bf{m}}\sigma'}a_{\ell,{\bf{m}}\sigma'}\right.\right.
\nonumber\\
\left.\left.+b^{\dag}_{\ell,{\bf{m}}\sigma'}b_{\ell,{\bf{m}}\sigma'}\right)\right]=-2\gamma_1\sum_{{\bf{m}}\sigma}{\bf{R}}_{m}b^{\dag}_{1{\bf{m}}\sigma}a_{2{\bf{m}}\sigma}.
\label{Equation_7_7} 
\end{eqnarray} 
The vanishing of this expression becomes apparent when passing into the Fourier space representation for the fermionic operators $b^{\dag}_{1{\bf{m}}\sigma}$ and $a_{2{\bf{m}}\sigma}$. We have
\begin{eqnarray}    
b^{\dag}_{1{\bf{m}}\sigma}=\frac{1}{\sqrt{N}}\sum_{{\bf{k}}}b^{\dag}_{1{\bf{k}}\sigma}e^{-i{\bf{k}}{\bf{R}}_{m}},
\nonumber\\
a_{2{\bf{m}}\sigma}=\frac{1}{\sqrt{N}}\sum_{{\bf{k}}}a_{2{\bf{k}}\sigma}e^{i{\bf{k}}{\bf{R}}_{m}},
\label{Equation_7_8} 
\end{eqnarray} 
and we obtain
\begin{eqnarray}    
\left[h_{{\bf{n}}}(\gamma_1),n_{{\bf{m}}}\right]=-\frac{\gamma_1}{N^{2}}\sum_{{\bf{m}}\sigma}\sum_{{\bf{k}},\Delta{\bf{k}}}{\bf{R}}_{m}b^{\dag}_{1{\bf{k}}\sigma}a_{2{\bf{k}}+\Delta{\bf{k}}\sigma}e^{i\Delta{\bf{k}}{\bf{R}}_{m}},
\nonumber\\
\label{Equation_7_9} 
\end{eqnarray} 
where $\Delta{\bf{k}}$ is the wave vector difference $\Delta{\bf{k}}={\bf{k}}'-{\bf{k}}$. 
The sum over the lattice vectors ${\bf{m}}$ in the expression in Eq.(\ref{Equation_7_9}) is 
\begin{eqnarray}    
&&\sum_{{\bf{m}}}{\bf{R}}_{m}e^{i\Delta{\bf{k}}{\bf{R}}_{m}}=\sum_{{{m}}_{1}{{m}}_{2}}\left({{m}}_{1}{\bf{a}}_{1}+{{m}}_{2}{\bf{a}}_{2}\right)e^{i\Delta{\bf{k}}\left({{m}}_{1}{\bf{a}}_{1}+{{m}}_{2}{\bf{a}}_{2}\right)}
\nonumber\\
&&=N\delta\left(\Delta{\bf{k}}\right)\left({\bf{a}}_{1}\sum_{m_{1}}m_{1}+{\bf{a}}_{2}\sum_{m_{2}}m_{2}\right)=0.
\label{Equation_7_10} 
\end{eqnarray} 
Here, ${\bf{a}}_{1}$ and ${\bf{a}}_{2}$ are the unit cell vectors defined in the 
Section \ref{sec:Section_3} and $m_{1}$ and $m_{2}$ are integer numbers. We have used, in Eq.(\ref{Equation_7_10}), the fact that $\sum_{{\bf{m}}}e^{i\Delta{\bf{k}}{\bf{R}}_{m}}=N\delta\left({\Delta{\bf{k}}}\right)$ and $\sum_{{m}_{i}}{m}_{i}=0$, because we consider the infinite lattice structure. Furthermore, we will use the expression of the electric current density operator, given in Eq.(\ref{Equation_7_6}) when evaluating the linear response coefficients, presented in the Section \ref{sec:Section_4}. The Fourier transformed form of the expression in Eq.(\ref{Equation_7_6}) could be done easier by using the rules in Eq.(\ref{Equation_8}). After a simple calculation we write
\begin{eqnarray}    
{\bf{j}}_{\rm e}=\frac{1}{N}\sum_{{\bf{k}}\sigma}\left(a^{\dag}_{1{\bf{k}}\sigma}b_{1{\bf{k}}\sigma}{\bf{v}}_{{\bf{k}}}+b^{\dag}_{2{\bf{k}}\sigma}a_{2{\bf{k}}\sigma}{\bf{v}}_{{\bf{k}}}+{\rm h.c.}\right),
\nonumber\\
\label{Equation_7_11} 
\end{eqnarray} 
where the ${\bf{k}}$-dependent velocity operator ${\bf{v}}_{{\bf{k}}}$ is defined in Eq.(\ref{Equation_31}). Furthermore, we can use the fermionic Nambu spinors notations, introduced in Section \ref{sec:Section_2_1}, and we write the electric current density operator in the more symmetric form. Namely, we have
\begin{eqnarray}    
{\bf{j}}_{\rm e}=\frac{1}{N}\sum_{{\bf{k}}\sigma}\Psi^{\dag}_{{\bf{k}}\sigma}(\nu_{n}){\upsilon}_{{\bf{k}}}\Psi_{{\bf{k}}\sigma}(\nu_{n}),
\label{Equation_7_12} 
\end{eqnarray}
where the electron velocity matrix ${\upsilon}_{\bf{k}}$ is  
\begin{eqnarray}
{\upsilon}_{{\bf{k}}}=\left(
\begin{array}{ccccrrrr}
0 & 0 & 0 & {\bf{v}}_{{\bf{k}}}\\
0 &0  & {\bf{v}}_{{\bf{k}}} & 0 \\
0 & {\bf{v}}^{\ast}_{{\bf{k}}} & 0 & 0 \\
{\bf{v}}^{\ast}_{{\bf{k}}} & 0 & 0 & 0
\end{array}
\right).
\label{Equation_7_13}
\end{eqnarray}
At the end, we remark that the commutation of the electronic density operator $n_{{\bf{m}}}$ with the other terms containing the single electron density operators is obvious and we don't give here those derivations. 
%
\subsection{\label{sec:Section_7_2} Total heat current density operator ${\bf{j}}_{\rm H}$}
%
We define an operator which is the sum over the position $R_{{\bf{m}}}$ and the effective Hamiltonian density operator $h_{\rm eff{\bf{m}}}$ given in Eq.(\ref{Equation_7_3}), i.e., ${\bf{R}}_{\rm E}=\sum_{{\bf{m}}}{\bf{R}}_{{\bf{m}}}h_{\rm eff{\bf{m}}}$. Next we use the equation of continuity $\partial{H}/\partial{\tau}-i\nabla{{\bf{j}}}_{\rm E}=0$ which leads to the following equation for the energy current density operator ${\bf{j}}_{\rm E}=i\partial{{\bf{R}}_{\rm E}}/\partial{\tau}$, i.e., ${\bf{j}}_{\rm E}=i\left[{\bf{R}}_{\rm E},H\right]$, where we have used the Heisenberg equation of motion for the operator ${\bf{R}}_{\rm E}$: $\partial{{\bf{R}}_{\rm E}}/\partial{\tau}=-\left[{\bf{R}}_{\rm E},H\right]$. Then we have  
\begin{eqnarray}    
{\bf{j}}_{\rm E}=i\sum_{{\bf{m}}{\bf{n}}}{\bf{R}}_{{\bf{m}}}\left[h_{\rm eff{\bf{n}}},h_{\rm eff {\bf{m}}}\right].
\label{Equation_7_14} 
\end{eqnarray}
Therefore, the calculation of the energy current density operator is reduced to the calculation of the commutator between the effective Hamiltonian densities at different site positions $h_{\rm eff{\bf{n}}}$ and $h_{\rm eff{\bf{m}}}$. When calculating this commutator with the effective Hamiltonian density given in Eq.(\ref{Equation_7_3}), we should take into account the commutator between (a) intralayer-intralayer hopping terms, (b) intralayer and interlayer hopping terms, (c) intralayer and effective chemical potential terms and finally (d) interlayer and chemical potential terms. The last commutator is zero as it was discussed above. We present here the calculations of the other commutators, cited here.  
%
\subsubsection{\label{sec:Section_7_2_1} Contribution of the intralayer hopping terms}
%
The commutator between the intralayer terms with the Hamiltonian densities $h_{\rm eff{\bf{n}}}(\gamma_0)$ and $h_{\rm eff{\bf{m}}}(\gamma_0)$ has the contribution 
\begin{eqnarray}    
&&i\sum_{{\bf{m}}{\bf{n}}}{\bf{R}}_{{\bf{m}}}\left[h_{\rm eff{\bf{n}}}(\gamma_0),h_{\rm eff{\bf{m}}}(\gamma_0)\right]
\nonumber\\
&&=i\gamma^{2}_{0}\sum_{{\bf{m}}{\bf{n}}}{\bf{R}}_{{\bf{m}}}\sum_{\substack{\ell\sigma\sigma'\\ {\bm{\mathit{\delta}}}_{\ell}{\bm{\mathit{\delta}}}'_{\ell}}}\left(a^{\dag}_{\ell,{\bf{n}}\sigma}a_{\ell,{\bf{m}}\sigma'}-{\rm h.c.}\right)\delta_{{\bf{n}}+{\bm{\mathit{\delta}}}_{\ell},{\bf{m}}+{\bm{\mathit{\delta}}}'_{\ell}}\delta_{\sigma\sigma'}
\nonumber\\
&&+\left(b^{\dag}_{\ell,{\bf{n}}+{\bm{\mathit{\delta}}}_{\ell}\sigma}b_{\ell,{\bf{m}}+{\bm{\mathit{\delta}}}'_{\ell}\sigma'}-{\rm h.c.}\right)\delta_{{\bf{n}},{\bf{m}}}\delta_{\sigma\sigma'}.
\nonumber\\
&&=i\gamma^{2}_{0}\sum_{{\bf{m}}\ell}\sum_{\substack{{\bm{\mathit{\delta}}}_{\ell}{\bm{\mathit{\delta}}}'_{\ell} \\ \sigma}}\left({\bm{\mathit{\delta}}}'_{\ell}-{\bm{\mathit{\delta}}}_{\ell}\right)a^{\dag}_{\ell,{\bf{m}}+{\bm{\mathit{\delta}}}'_{\ell}-{\bm{\mathit{\delta}}}_{\ell},\sigma}a_{\ell, {\bf{m}}\sigma}.
\label{Equation_7_15} 
\end{eqnarray}
Furthermore, we can use the transformations in Eq.(\ref{Equation_7_8}) and we have
\begin{eqnarray}    
&&i\sum_{{\bf{m}}{\bf{n}}}{\bf{R}}_{{\bf{m}}}\left[h_{\rm eff{\bf{n}}}(\gamma_0),h_{\rm eff{\bf{m}}}(\gamma_0)\right]=
\nonumber\\
&&=i\frac{\gamma^{2}_{0}}{N}\sum_{\substack{\ell\sigma \\ {\bm{\mathit{\delta}}}_{\ell}{\bm{\mathit{\delta}}}'_{\ell}}}\sum_{{\bf{k}}}\left({\bm{\mathit{\delta}}}'_{\ell}-{\bm{\mathit{\delta}}}_{\ell}\right)a^{\dag}_{\ell,{\bf{k}}\sigma}a_{\ell,{\bf{k}}\sigma}e^{-i{\bf{k}}\left({\bm{\mathit{\delta}}}'_{\ell}-{\bm{\mathit{\delta}}}_{\ell}\right)}.
\nonumber\\
\label{Equation_7_16} 
\end{eqnarray}
It is easy to realize that 
\begin{eqnarray}    
\sum_{{\bm{\mathit{\delta}}}_{\ell}{\bm{\mathit{\delta}}}'_{\ell}}\left({\bm{\mathit{\delta}}}'_{\ell}-{\bm{\mathit{\delta}}}_{\ell}\right)e^{-i{\bf{k}}\left({\bm{\mathit{\delta}}}'_{\ell}-{\bm{\mathit{\delta}}}_{\ell}\right)}=\sum_{\substack{{\bf{u}}=0, \pm {\bf{a}}_{1}\\ \pm{\bf{a}}_{2}}}{\bf{u}}e^{-i{\bf{k}}{\bf{u}}},
\label{Equation_7_17} 
\end{eqnarray}
where the vectors ${\bf{a}}_{1}$ and ${\bf{a}}_{2}$ in the summation in the right side in Eq.(\ref{Equation_7_17}) are the unit cell vectors. The relation in Eq.(\ref{Equation_7_17}) is true for both layers $\ell=1,2$ because of the obvious relation ${\bm{\mathit{\delta}}}'_{2}-{\bm{\mathit{\delta}}}_{2}={\bm{\mathit{\delta}}}_{1}-{\bm{\mathit{\delta}}}'_{1}$. 
%
\subsubsection{\label{sec:Section_7_2_2} Contribution of the interlayer hopping term}
%
\begin{eqnarray}  
&&i\sum_{{\bf{m}}{\bf{n}}}{\bf{R}}_{{\bf{m}}}\left[h_{\rm eff{\bf{n}}}(\gamma_0),h_{\rm eff{\bf{m}}}(\gamma_1)\right]
\nonumber\\
&&=i\gamma_0\left(\gamma_1+\Delta\right)\sum_{\substack{{\bf{m}}{\bm{\mathit{\delta}}}_{1}\\ \sigma}}{\bf{R}}_{{\bf{m}}+{\bm{\mathit{\delta}}}_{1}}\left(a^{\dag}_{1,{\bf{m}}\sigma}a_{2,{\bf{m}}+{\bm{\mathit{\delta}}}_{1}\sigma}-{\rm h.c.}\right)
\nonumber\\
&&+i\gamma_0\left(\gamma_1+\Delta\right)\sum_{\substack{{\bf{m}}{\bm{\mathit{\delta}}}_{2}\\ \sigma}}{\bf{R}}_{{\bf{m}}-{\bm{\mathit{\delta}}}_{2}}\left(b^{\dag}_{2,{\bf{m}}\sigma}b_{1,{\bf{m}}-{\bm{\mathit{\delta}}}_{2}\sigma}-{\rm h.c.}\right)
\nonumber\\
&&=i\frac{\gamma_0\left(\gamma_1+\Delta\right)}{N}\sum_{{\bf{k}}\sigma}\left[a^{\dag}_{1,{\bf{k}}\sigma}a_{2,{\bf{k}}\sigma}\sum_{{\bm{\mathit{\delta}}}_{1}}\left({\bm{\mathit{\delta}}}_{1}e^{i{\bf{k}}{\bm{\mathit{\delta}}}_{1}}\right)-{\rm h.c.}\right.
\nonumber\\
&&\left.+b^{\dag}_{2,{\bf{k}}\sigma}b_{1,{\bf{k}}\sigma}\sum_{{\bm{\mathit{\delta}}}_{1}}\left({\bm{\mathit{\delta}}}_{1}e^{i{\bf{k}}{\bm{\mathit{\delta}}}_{1}}\right)-{\rm h.c.}
\right].
\label{Equation_7_18} 
\end{eqnarray}
Here, we have used the fact that ${\bm{\mathit{\delta}}}_{2}=-{\bm{\mathit{\delta}}}_{1}$. Let's remark also that
\begin{eqnarray}    
\sum_{{\bm{\mathit{\delta}}}_{1}}{\bm{{\delta}}}_{1}e^{i{\bf{k}}{\bm{\mathit{\delta}}}_{1}}=\sum_{\substack{{\bf{a}}'=0, {\bf{a}}_{1},{\bf{a}}_{2}}}{\bf{a}}'e^{i{\bf{k}}{\bf{a}}'}.
\label{Equation_7_19} 
\end{eqnarray}
Thus the summation over ${\bm{\mathit{\delta}}}_{1}$ is equivalent to the summation over the lattice vectors ${\bf{a}}'=0, {\bf{a}}_{1}, {\bf{a}}_{2}$. Therefore, we can write
\begin{eqnarray}  
&&i\sum_{{\bf{m}}{\bf{n}}}{\bf{R}}_{{\bf{m}}}\left[h_{\rm eff{\bf{n}}}(\gamma_0),h_{\rm eff{\bf{m}}}(\gamma_1)\right]=
\nonumber\\
&&=i\frac{\gamma_0\left(\gamma_1+\Delta\right)}{N}\sum_{{\bf{k}}\sigma}\sum_{{\bf{a}}'}{\bf{a}}'\left(a^{\dag}_{1,{\bf{k}}\sigma}a_{2,{\bf{k}}\sigma}e^{i{\bf{k}}{{\bf{a}}'}}\right.
\nonumber\\
&&\left. +b^{\dag}_{2,{\bf{k}}\sigma}b_{1,{\bf{k}}\sigma}e^{i{\bf{k}}{{\bf{a}}}'}-{\rm h.c.}
\right).
\label{Equation_7_20} 
\end{eqnarray}
%
\subsubsection{\label{sec:Section_7_2_3} Effective chemical potential terms}
%
Here, we present the calculation of the commutator between the intralayer hopping term and the last two terms in the effective Hamiltonian density given in Eq.(\ref{Equation_7_3}). The commutator which contains the effective chemical potential $\mu_{\rm 1eff}$ is 
\begin{eqnarray}  
&&i\sum_{{\bf{m}}{\bf{n}}}{\bf{R}}_{{\bf{m}}}\left[h_{\rm eff{\bf{n}}}(\gamma_0),h_{\rm eff{\bf{m}}}(\mu_{\rm 1eff})\right]
\nonumber\\
&&=i\mu_{\rm 1eff}\gamma_{0}\sum_{{\bf{m}}{\bm{{\delta}}}_{1}\sigma}{\bf{R}}_{{\bf{m}}}\left(b^{\dag}_{1,{\bf{m}}+{\bm{{\delta}}}_{1}}a_{1,{\bf{m}}\sigma}-{\rm h.c.}\right)
\nonumber\\
&&+i\mu_{\rm 1eff}\gamma_{0}\sum_{{\bf{m}}{\bm{{\delta}}}_{2}\sigma}{\bf{R}}_{{\bf{m}}+{\bm{{\delta}}}_{2}}\left(a^{\dag}_{2,{\bf{m}}}b_{2,{\bf{m}}+{\bm{{\delta}}}_{2}}-{\rm h.c.}\right)
\nonumber\\
&&=i\mu_{\rm 1eff}\gamma_{0}\sum_{{\bf{m}}{\bm{{\delta}}}_{2}\sigma}{\bm{{\delta}}}_{2}\left(a^{\dag}_{2,{\bf{m}}}b_{2,{\bf{m}}+{\bm{{\delta}}}_{2}}-{\rm h.c.}\right)
\label{Equation_7_21} 
\end{eqnarray}
or, transforming it into the Fourier space we can write
\begin{eqnarray}  
&&i\sum_{{\bf{m}}{\bf{n}}}{\bf{R}}_{{\bf{m}}}\left[h_{\rm eff{\bf{n}}}(\gamma_0),h_{\rm eff{\bf{m}}}(\mu_{\rm 1eff})\right]=
\nonumber\\
&&=i\frac{\mu_{\rm 1eff}\gamma_{0}}{N}\sum_{{\bf{k}}{\bm{{\delta}}}_{2}\sigma}{\bm{{\delta}}}_{2}\left(a^{\dag}_{2,{\bf{k}}}b_{2,{\bf{k}}}e^{i{\bf{k}}{\bm{{\delta}}}_{2}}-{\rm h.c.}\right).
\label{Equation_7_22} 
\end{eqnarray}
Similarly, the commutator which contains the effective chemical potential $\mu_{\rm 2eff}$ is
\begin{eqnarray}  
&&i\sum_{{\bf{m}}{\bf{n}}}{\bf{R}}_{{\bf{m}}}\left[h_{\rm eff{\bf{n}}}(\gamma_0),h_{\rm eff{\bf{m}}}(\mu_{\rm 2eff})\right]
\nonumber\\
&&=i\mu_{\rm 2eff}\gamma_{0}\sum_{{\bf{m}}{\bm{{\delta}}}_{2}\sigma}{\bf{R}}_{{\bf{m}}}\left(b^{\dag}_{2,{\bf{m}}+{\bm{{\delta}}}_{2}}a_{2,{\bf{m}}\sigma}-{\rm h.c.}\right)
\nonumber\\
&&+i\mu_{\rm 2eff}\gamma_{0}\sum_{{\bf{m}}{\bm{{\delta}}}_{2}\sigma}{\bf{R}}_{{\bf{m}}+{\bm{{\delta}}}_{1}}\left(a^{\dag}_{1,{\bf{m}}}b_{1,{\bf{m}}+{\bm{{\delta}}}_{1}}-{\rm h.c.}\right)
\nonumber\\
&&=i\mu_{\rm 2eff}\gamma_{0}\sum_{{\bf{m}}{\bm{{\delta}}}_{1}\sigma}{\bm{{\delta}}}_{1}\left(a^{\dag}_{1,{\bf{m}}}b_{1,{\bf{m}}+{\bm{{\delta}}}_{1}}-{\rm h.c.}\right).
\label{Equation_7_23} 
\end{eqnarray}
The transformation into the Fourier space gives the following result
\begin{eqnarray}  
&&i\sum_{{\bf{m}}{\bf{n}}}{\bf{R}}_{{\bf{m}}}\left[h_{\rm eff{\bf{n}}}(\gamma_0),h_{\rm eff{\bf{m}}}(\mu_{\rm 2eff})\right]=
\nonumber\\
&&=i\frac{\mu_{\rm 2eff}\gamma_{0}}{N}\sum_{{\bf{k}}{\bm{{\delta}}}_{1}\sigma}{\bm{{\delta}}}_{1}\left(a^{\dag}_{1,{\bf{k}}}b_{1,{\bf{k}}}e^{i{\bf{k}}{\bm{{\delta}}}_{1}}-{\rm h.c.}\right).
\label{Equation_7_24} 
\end{eqnarray}
At the end, we can combine the expressions obtained in Eqs.(\ref{Equation_7_16}), (\ref{Equation_7_20}), (\ref{Equation_7_22}) and (\ref{Equation_7_24}) and write the expression of the total energy current density operator ${\bf{j}}_{\rm E}$ in the Fourier space. We get 
\begin{eqnarray}  
&&{\bf{j}}_{\rm E}=i\frac{\gamma^{2}_{0}}{N}\sum_{\ell=1,2}\sum_{{\bf{k}}\sigma}\sum_{\substack{{\bf{u}}=0, \pm {\bf{a}}_{1}\\ \pm{\bf{a}}_{2}}}{\bf{u}}e^{-i{\bf{k}}{\bf{u}}}a^{\dag}_{\ell,{\bf{k}}\sigma}a_{\ell,{\bf{k}}\sigma}
\nonumber\\
&&+i\frac{\gamma_0\left(\gamma_1+\Delta\right)}{N}\sum_{{\bf{k}}\sigma}\sum_{{\bf{a}}'}{\bf{a}}'\left(a^{\dag}_{1,{\bf{k}}\sigma}a_{2,{\bf{k}}\sigma}e^{i{\bf{k}}{\bf{a}}'}+b^{\dag}_{2,{\bf{k}}\sigma}b_{1,{\bf{k}}\sigma}e^{i{\bf{k}}{\bf{a}}'}\right.
\nonumber\\
&&\left.-{\rm h.c.}
\right)
\nonumber\\
&&+i\frac{\mu_{\rm 1eff}\gamma_{0}}{N}\sum_{{\bf{k}}{\bm{{\delta}}}_{2}\sigma}{\bm{{\delta}}}_{2}\left(a^{\dag}_{2,{\bf{k}}}b_{2,{\bf{k}}}e^{i{\bf{k}}{\bm{{\delta}}}_{2}}-{\rm h.c.}\right)
\nonumber\\
&&
+i\frac{\mu_{\rm 2eff}\gamma_{0}}{N}\sum_{{\bf{k}}{\bm{{\delta}}}_{1}\sigma}{\bm{{\delta}}}_{1}\left(a^{\dag}_{1,{\bf{k}}}b_{1,{\bf{k}}}e^{i{\bf{k}}{\bm{{\delta}}}_{1}}-{\rm h.c.}\right).
\label{Equation_7_25} 
\end{eqnarray}
Next, the total heat current density operator ${\bf{j}}_{\rm H}$ is defined in Eq.(\ref{Equation_22}) in the Section \ref{sec:Section_3}. Taking into account the expression of the electric current operator, given in Eq.(\ref{Equation_7_11}) we can write for ${\bf{j}}_{\rm H}$
\begin{eqnarray}  
&&{\bf{j}}_{\rm H}={\bf{j}}_{\rm E}-\mu{{\bf{j}}_{\rm e}}=\frac{1}{N}\sum_{{\bf{k}}\sigma}\sum_{\ell}a^{\dag}_{\ell,{\bf{k}}\sigma}a_{\ell,{\bf{k}}\sigma}{\bf{v}}_{1{\bf{k}}}
\nonumber\\
&&+\frac{1}{N}\sum_{{\bf{k}}\sigma}\left[\left(a^{\dag}_{1,{\bf{k}}\sigma}a_{2,{\bf{k}}\sigma}+b^{\dag}_{2,{\bf{k}}\sigma}b_{1,{\bf{k}}\sigma}\right){\bf{v}}_{2{\bf{k}}}\right.
\nonumber\\
&&\left.+b^{\dag}_{2,{\bf{k}}\sigma}a_{2,{\bf{k}}\sigma}{\bf{v}}_{3{\bf{k}}} +a^{\dag}_{1,{\bf{k}}\sigma}b_{1,{\bf{k}}\sigma}{\bf{v}}_{4{\bf{k}}}+{\rm h.c.}\right],
\label{Equation_7_26} 
\end{eqnarray}
where the velocity operators ${\bf{v}}_{i{\bf{k}}}$ with $i=1,...4$ are defined in Eq.(\ref{Equation_36}), in the Section \ref{sec:Section_3}. After using the Nambu spinor representation for the fermionic operators, the result in Eq.(\ref{Equation_7_26}) can be rewritten in the form given in Eq.(\ref{Equation_32}), in the Section \ref{sec:Section_3}.
%
\section{\label{sec:Section_8} The linear response function ${\cal{L}}_{\rm 22}(\omega_{m})$}
%
We give here the form of the heat-heat response function in terms of the normal and excitonic Green's functions defined in the Section \ref{sec:Section_2_1}. The function ${\cal{L}}_{\rm 22}(\omega_{m})$ is defined in the Kubo-Green theory as  
\begin{eqnarray}    
{\cal{L}}_{\rm 22}(\omega_{m})=\frac{1}{\beta{\omega_{m}}}\int^{\beta}_{0}d\tau{e^{i\omega_{m}\tau}}\left\langle T_{\tau}{\bf{j}}_{\rm H}(\tau){\bf{j}}_{\rm H}(0)\right\rangle.
\label{Equation_8_1}
\end{eqnarray}
For calculating the correlator under the integral in the right hand side in Eq.(\ref{Equation_8_1}) we use the form of the total heat current density operator, given in Eq.(\ref{Equation_7_26}), in the Section \ref{sec:Section_7_2_3}. When expanding the correlator we get $100$ terms from which survive only $18$, because many of them vanish when expanding the four fermion correlators with the help of the Wick theorem. As an example, we consider a term which gives the product of the normal Green's function with the single excitonic correlator  
\begin{eqnarray}    
&&\left\langle a^{\dag}_{2,{\bf{k}}\sigma}(\tau)b_{2{\bf{k}}\sigma}(\tau)b^{\dag}_{2,{\bf{k}}'\sigma'}(0)b_{1,{\bf{k}}'\sigma'}(0) \right\rangle=
\nonumber\\
&&=\left\langle a^{\dag}_{2,{\bf{k}}\sigma}(\tau)b_{2{\bf{k}}\sigma}(\tau)\right\rangle \left\langle b^{\dag}_{2,{\bf{k}}'\sigma'}(0)b_{1,{\bf{k}}'\sigma'}(0)\right\rangle
\nonumber\\
&&-\left\langle a^{\dag}_{2,{\bf{k}}\sigma}(\tau)b^{\dag}_{2{\bf{k}}'\sigma'}(0)\right\rangle \left\langle b_{2,{\bf{k}}\sigma}(\tau)b_{1,{\bf{k}}'\sigma'}(0)\right\rangle
\nonumber\\
&&+\left\langle a^{\dag}_{2,{\bf{k}}\sigma}(\tau)b_{1{\bf{k}}'\sigma'}(0)\right\rangle 
\left\langle b_{2,{\bf{k}}\sigma}(\tau)b^{\dag}_{2,{\bf{k}}'\sigma'}(0)\right\rangle
\nonumber\\
&&=\left\langle a^{\dag}_{2,{\bf{k}}\sigma}(\tau)b_{1{\bf{k}}'\sigma'}(0)\right\rangle \left\langle b_{2,{\bf{k}}\sigma}(\tau)b^{\dag}_{2,{\bf{k}}'\sigma'}(0)\right\rangle.
\label{Equation_8_2}
\end{eqnarray}
The first two terms in Eq.(\ref{Equation_8_2}) vanish because of the symmetry of the action of the system. Then we have ${\cal{F}}_{a_{2}b_{1}}={\cal{F}}_{b_{1}a_{2}}$, and we get
\begin{eqnarray}    
&&\left\langle a^{\dag}_{2,{\bf{k}}\sigma}(\tau)b_{2{\bf{k}}\sigma}(\tau)b^{\dag}_{2,{\bf{k}}'\sigma'}(0)b_{1,{\bf{k}}'\sigma'}(0) \right\rangle=
\nonumber\\
&&=-{\cal{F}}_{b_{1}a_{2},{\bf{k}}'-{\bf{k}}}\left(-\tau\right){\cal{G}}_{b_{2},{\bf{k}}-{\bf{k}}'}\left(\tau\right).
\label{Equation_8_3}
\end{eqnarray}
Here, we give the final result for the function in Eq.(\ref{Equation_8_1}) 
\begin{eqnarray}    
&&{\cal{L}}_{\rm 22}(\omega_{m})=-\frac{1}{\beta^{2}{\omega_{m}}}\sum_{\substack{{\bf{k}}\sigma \\ \ell \nu_{n}}}{\cal{G}}^{\sigma}_{a_{\ell}{\bf{k}}}(\nu_{n}){\cal{G}}^{\sigma}_{a_{\ell}{\bf{k}}}(\nu_{n}+\omega_{m})|{\bf{v}}_{1{\bf{k}}}|^{2}
\nonumber\\
&&-\frac{1}{\beta^{2}{\omega_{m}}}\sum_{\substack{{\bf{k}}\sigma \\ \nu_{n}}}\left({\cal{G}}^{\sigma}_{a_{1}{\bf{k}}}(\nu_{n}){\cal{G}}^{\sigma}_{a_{2}{\bf{k}}}(\nu_{n}+\omega_{m})\right.
\nonumber\\
&&\left.+{\cal{G}}^{\sigma}_{a_{1}{\bf{k}}}(\nu_{n}){\cal{G}}^{\sigma}_{a_{2}{\bf{k}}}(\nu_{n}-\omega_{m})\right)|{\bf{v}}_{2{\bf{k}}}|^{2}
\nonumber\\
&&-\frac{1}{\beta^{2}{\omega_{m}}}\sum_{\substack{{\bf{k}}\sigma \\ \nu_{n}}}\left({\cal{G}}^{\sigma}_{b_{1}{\bf{k}}}(\nu_{n}){\cal{G}}^{\sigma}_{b_{2}{\bf{k}}}(\nu_{n}+\omega_{m})\right.
\nonumber\\
&&\left.+{\cal{G}}^{\sigma}_{b_{1}{\bf{k}}}(\nu_{n}){\cal{G}}^{\sigma}_{b_{2}{\bf{k}}}(\nu_{n}-\omega_{m})\right)|{\bf{v}}_{2{\bf{k}}}|^{2}
\nonumber\\
&&-\frac{1}{\beta^{2}{\omega_{m}}}\sum_{\substack{{\bf{k}}\sigma \\ \nu_{n}}}\left({\cal{G}}^{\sigma}_{a_{2}{\bf{k}}}(\nu_{n}){\cal{G}}^{\sigma}_{b_{2}{\bf{k}}}(\nu_{n}+\omega_{m})\right.
\nonumber\\
&&\left.+{\cal{G}}^{\sigma}_{a_{2}{\bf{k}}}(\nu_{n}){\cal{G}}^{\sigma}_{b_{2}{\bf{k}}}(\nu_{n}-\omega_{m})\right)|{\bf{v}}_{3{\bf{k}}}|^{2}
\nonumber\\
&&-\frac{1}{\beta^{2}{\omega_{m}}}\sum_{\substack{{\bf{k}}\sigma \\ \nu_{n}}}\left({\cal{G}}^{\sigma}_{a_{1}{\bf{k}}}(\nu_{n}){\cal{G}}^{\sigma}_{b_{1}{\bf{k}}}(\nu_{n}+\omega_{m})\right.
\nonumber\\
&&\left.+{\cal{G}}^{\sigma}_{a_{1}{\bf{k}}}(\nu_{n}){\cal{G}}^{\sigma}_{b_{1}{\bf{k}}}(\nu_{n}-\omega_{m})\right)|{\bf{v}}_{4{\bf{k}}}|^{2}
\nonumber\\
&&-\frac{1}{\beta^{2}{\omega_{m}}}\sum_{\substack{{\bf{k}}\sigma \\ \nu_{n}}}\left({\cal{G}}^{\sigma}_{a_{1}{\bf{k}}}(\nu_{n}){\cal{F}}^{\sigma}_{a_{2}b_{1}{\bf{k}}}(\nu_{n}+\omega_{m})\right.
\nonumber\\
&&\left.+{\cal{G}}^{\sigma}_{a_{1}{\bf{k}}}(\nu_{n}){\cal{F}}^{\sigma}_{a_{2}b_{1}{\bf{k}}}(\nu_{n}-\omega_{m})\right)\left({\bf{v}}_{2{\bf{k}}}{\bf{v}}^{\ast}_{4{\bf{k}}}+{\bf{v}}^{\ast}_{2{\bf{k}}}{\bf{v}}_{4{\bf{k}}}\right)
\nonumber\\
&&-\frac{1}{\beta^{2}{\omega_{m}}}\sum_{\substack{{\bf{k}}\sigma \\ \nu_{n}}}\left({\cal{F}}^{\sigma}_{a_{2}b_{1}{\bf{k}}}(\nu_{n}){\cal{G}}^{\sigma}_{b_{2}{\bf{k}}}(\nu_{n}+\omega_{m})\right.
\nonumber\\
&&\left.+{\cal{F}}^{\sigma}_{a_{2}b_{1}{\bf{k}}}(\nu_{n}){\cal{G}}^{\sigma}_{b_{2}{\bf{k}}}(\nu_{n}-\omega_{m})\right)\left({\bf{v}}_{2{\bf{k}}}{\bf{v}}^{\ast}_{3{\bf{k}}}+{\bf{v}}^{\ast}_{3{\bf{k}}}{\bf{v}}_{2{\bf{k}}}\right).
\label{Equation_8_4}
\end{eqnarray}
Furthermore, we use the explicit forms of the normal and excitonic Green's functions given in Eqs.(\ref{Equation_14}) and (\ref{Equation_17}) and we perform the summation over the fermionic Matsubara frequencies $\nu_{n}$. Then we use the Eq.(\ref{Equation_27}) and we get the expression of the response coefficient, given in Eq.(\ref{Equation_43}), in the Section \ref{sec:Section_4}. 

\end{document}